\documentclass[prl,letterpaper,english,reprint,nofootinbib,aps,superscriptaddress,showpacs,showkeys]{revtex4-1}
\usepackage{babel,calc,amsmath,amsthm,amssymb,graphicx,subfigure,xcolor,soul,bm}
\usepackage[T1]{fontenc}
\usepackage[unicode=true]{hyperref}
\hypersetup{
     colorlinks=true,       		
     linkcolor=blue,          	
     citecolor=blue,            
     urlcolor=blue,           	
}
\newtheorem{theorem}{Observation}
\newtheorem*{theorem*}{Theorem}

\newtheorem*{corollary*}{Corollary}

\newtheorem*{lemma*}{Lemma}

\newtheorem*{proposition*}{Proposition}
\theoremstyle{definition}

\newtheorem*{definition*}{Definition}
\theoremstyle{remark}

\newtheorem*{remark*}{Remark}

\newcommand{\ket}[1]{\left|#1\right\rangle}
\newcommand{\exv}[1]{\left\langle#1\right\rangle}

\newcommand{\eq}[1]{Eq.~\eqref{#1}}

\begin{document}
   \renewcommand{\figurename}{Fig.}
   \title{Generalized Iterative Formula for Bell Inequalities}

   \author{Xing-Yan Fan}
   \affiliation{Theoretical Physics Division, Chern Institute of Mathematics, Nankai University, Tianjin 300071, People's Republic
      of China}

\author{Zhen-Peng Xu}
    \email{zhen-peng.xu@ahu.edu.cn}
    \affiliation{School of Physics and Optoelectronics Engineering, Anhui University, 230601 Hefei, People’s Republic of China}

   \author{Jia-Le Miao}
   \affiliation{CAS Key Laboratory of Quantum Information, University of Science and Technology of China, Hefei, 230026, People's Republic of China}

   \author{Hong-Ye Liu}
   \affiliation{CAS Key Laboratory of Theoretical Physics, Institute of Theoretical Physics, Chinese Academy of Sciences, Beijing 100190, China}

   \author{Yi-Jia Liu}
   \affiliation{CAS Key Laboratory of Quantum Information, University of Science and Technology of China, Hefei, 230026, People's Republic of China}

   \author{Wei-Min Shang}
   \affiliation{School of Science, Tianjin Chengjian University, Tianjin 300384, People's Republic of China}
   \affiliation{Theoretical Physics Division, Chern Institute of Mathematics, Nankai University, Tianjin 300071, People's Republic
      of China}

   \author{Jie Zhou}
   \affiliation{Theoretical Physics Division, Chern Institute of Mathematics, Nankai University, Tianjin 300071, People's Republic
      of China}

   \author{Hui-Xian Meng}
   \affiliation{School of Mathematics and Physics, North China Electric Power University, Beijing 102206, People's Republic of
      China}

   \author{Otfried G\"{u}hne}
   \email{otfried.guehne@uni-siegen.de}
   \affiliation{Naturwissenschaftlich-Technische Fakult\"at, Universit\"at Siegen, Walter-Flex-Stra\ss e 3, 57068 Siegen,Germany}

   \author{Jing-Ling~Chen}
   \email{chenjl@nankai.edu.cn}
   \affiliation{Theoretical Physics Division, Chern Institute of Mathematics, Nankai University, Tianjin 300071, People's Republic
      of China}

   \date{\today}
   \begin{abstract}
         Bell inequalities are a vital tool to detect the nonlocal correlations, but the construction of them for multipartite systems is still a complicated problem. In this work, inspired via a decomposition of $(n+1)$-partite Bell inequalities into $n$-partite ones, we present a generalized iterative formula to construct nontrivial $(n+1)$-partite ones from the $n$-partite ones. Our iterative formulas recover the well-known Mermin-Ardehali-Belinski{\u{\i}}-Klyshko (MABK) and other  families in the literature as special cases. Moreover, a family of ``dual-use'' Bell inequalities is proposed, in the sense that for the generalized Greenberger-Horne-Zeilinger states these inequalities lead to the same quantum violation as the MABK family and, at the same time, the inequalities are able to detect the non-locality in the entire entangled region. Furthermore, we present generalizations of the the I3322 inequality 
         to any $n$-partite case which are still tight, and of the $46$ \'{S}liwa's inequalities to the four-partite tight ones, by applying our iteration method to each inequality and its equivalence class. 
   \end{abstract}
   \maketitle

\emph{Introduction.}---
    The fact that quantum mechanical correlations are incompatible with local hidden variable (LHV) models \cite{1966Bell} is known as Bell nonlocality \cite{2014RMPBN}, which is usually revealed by the violation of Bell inequalities \cite{1964Bell}. As an extraordinary nonclassical phenomenon, Bell nonlocality plays an important role both on the theoretical level and in applications \cite{2010Buhrman}. On the one hand, Bell nonlocality is related to many other quantum correlations like quantum entanglement \cite{2009RMPHorodecki}, quantum steering \cite{2020Uola} and quantum contextuality \cite{2021Cabelloa}. On the other hand, Bell nonlocality is crucial in quantum applications like quantum key distribution \cite{2021QErik}, quantum randomness \cite{2013Dhara,2016Acin} and quantum computation \cite{2012Howard}.

    Regarding the importance of Bell nonlocality, the construction of Bell inequalities is always one central topic in quantum information theory. The first Bell inequality was discovered by Bell in a bipartite scenario to tackle the long-standing paradox proposed by Einstein, Podolsky and Rosen~\cite{1935EPR} from the statistical aspect. However, the original Bell inequality is hard to test experimentally. A revised version of the original Bell inequality, the Clause-Horne-Shimony-Holt (CHSH) inequality \cite{1969CHSH}, is more friendly for experiments. Employing the CHSH inequality, three loophole-free Bell experiments succeeded finally \cite{2015PRLMarissa,2015NHensen,2015PRLShalm}. The development of Bell inequalities has many directions, for example, Bell inequalities which are more robust against imperfectness in experiments \cite{2002Massar,2022Meng}, and the generalization of Bell inequalities into multipartite scenarios, see e.g., \cite{2002Zukowski,2003Sliwa,2013Wu}.

    A given LHV model can be described by a polytope, whose facets define the tight Bell inequalities
    \cite{1999Peres,2005Pironio,2006Avis}. In turn, all the tight Bell inequalities characterize the corresponding LHV model completely. However, the full characterization is usually a hard problem. This has only been done for simple $(n, m, d)$-scenarios, i.e., the $(2,2,2)$ \cite{1969CHSH}, $(2,2,3)$ \cite{2002Collinsa}, $(2,3,2)$ \cite{1981Froissart} and $(3,2,2)$ \cite{2003Sliwa} scenarios, where $n, m$ and $d$ stand for the number of parties, measurements per party and outcomes per measurement, respectively. For more complicated scenarios, all Bell inequalities can sometimes be characterized, if additional symmetries are
    considered~\cite{bancal2010looking,bernards2020generalizing}. Iterative formulas have then been a powerful alternative approach to find Bell inequalities of interest, especially in the multipartite case \cite{2001PRAWernerWolf,1990PRLM,1992PRAMA,1993PUBK,2002Collins,2002Seevinck,2006CAF}. Mermin \emph{et al.} presented an iteration method, which has successfully been generalized the CHSH inequality to the Mermin-Ardehali-Belinski{\u{\i}}-Klyshko (MABK) inequalities for arbitrary $n$-partite scenario~\cite{1990PRLM,1992PRAMA,1993PUBK}. As it turned out, the MABK inequalities are the strongest ones to detect the multi-qubit Greenberger-Horne-Zeilinger (GHZ) state \cite{1989GHZ}, in the sense that the quantum-classical ratio is the highest \cite{2001PRAWernerWolf,2004PRLZqC}. To detect genuine multipartite entanglement and nonlocality, Svetlichny constructed a tripartite inequality \cite{1987Svetlichny}, which was then generalized to the  multipartite scenario through iteration~\cite{2002Collins,2002Seevinck}. Chen-Albeverio-Fei (CAF) inequalities~\cite{2006CAF} are other relevant examples, which can probe the generalized GHZ state $\cos\theta\ket{000}+\sin\theta\ket{111}$ in the whole entangled region.

    However, all those iterative formulas are very specific and shed little light on new constructions. In this work, we firstly observe that any Bell inequality in the $(n+1, m, 2)$ scenario always has a decomposition into $n$-partite ones, which can be obtained by enumerating all the extremal classical assignments for the measurements of the last party. Based on this observation, we propose a generalized iterative formula to construct nontrivial $(n+1)$-partite Bell inequalities from a set of $n$-partite ones. Our iterative formula can recover several families in the literature as special cases, like the  MABK inequalities and CAF inequalities. Moreover, we come up with new families of Bell inequalities for specific purposes. For example, a family of ``dual-use'' inequalities is proposed. On the one hand, those inequalities have the quantum violation for the full entangled range of the generalized GHZ state as the CAF inequalities. On the other hand, the maximal quantum violation is as large as the one for MABK inequalities. Furthermore, we simplify our iterative formula by introducing more linear or symmetric constraints among the $n$-partite Bell inequalities. The $46$ \'Sliwa's inequalities \cite{2003Sliwa} for the $(3,2,2)$ scenario and the I3322 inequality \cite{1981Froissart} for the $(2,3,2)$ scenario are taken as examples.

\emph{Decomposition of Bell Inequalities and the generalized iterative formula}.---
    Firstly, we introduce the notation of Bell function, e.g., any $(n+1)$-partite Bell function can be written as $\mathbb{B}_{n+1} =\sum_{i\cdots jk} \alpha_{i\cdots jk} A_i \cdots B_j C_k$, where $A_i, B_j, C_k$ are random variables for different parties. Then $\exv{\mathbb{B}_{n+1}} \leq l$ defines one Bell inequality, where $l$ is the maximal expectation $\exv{\mathbb{B}_{n+1}}_{\rm max}$ for any probability distribution coming from a LHV model. Furthermore, we call a Bell inequality normalized, if its LHV upper bound equals $l = 1$. Since $\exv{\mathbb{B}_{n+1}}$ is a linear function of $\mathbb{B}_{n+1}$ either in the LHV model or in quantum theory, and we only care about the maximum of $\exv{\mathbb{B}_{n+1}}$ in each case, we do not distinguish $\exv{\mathbb{B}_{n+1}}$ and $\mathbb{B}_{n+1}$ in the following, if there is no risk of confusion. Then we can formulate:
    \begin{theorem}\label{observation1}
        Consider a Bell function 
        $
            \mathbb{B}_{n+1} =\sum_{i\cdots jk} \alpha_{i\cdots jk} A_i \cdots B_j C_k   
        $ 
        in the $(n+1, m, 2)$ scenario. 
        Then $ \mathbb{B}_{n+1} =\sum_{\bm s} \mathbb{B}_{n}^{\bm s} \prod_k \left[(1+s_k C_k)/2\right]$, where  $s_k \in \{-1, 1\}$, ${\bm s} = (s_1, \ldots, s_m)$ denotes a vector labeling all possible measurement results for the last party and  consequently $\mathbb{B}_{n}^{\bm s} =\sum_{i\cdots jk} \alpha_{i\cdots jk} \,s_k A_i \cdots B_j$.
    \end{theorem}
    \begin{proof}
        The direct expansion shows that
        \begin{align}\label{eq:expansion}
            & \sum_{\bm s} s_k \prod_{k'} \Big[\frac{(1+s_{k'} C_{k'})}{2}\Big]\notag\\
            =& \sum_{\bm s} \Big[\Big(\dfrac{s_k +C_k}{2}\Big)\prod_{k'\neq k} \Big(\dfrac{1+s_{k'} C_{k'}}{2}\Big)\Big] 
            = C_k.
        \end{align}
        The last equality follows from the fact that 
        $\sum_{s_k} [(s_k +C_k)/2] = C_k$ and $\sum_{s_{k'}} [(1 + s_{k'}C_{k'})/2]=1$ for each $k' \neq k$.
        
        By changing the order of summation, we have
        \begin{align}\label{eq:summation}
            &\sum_{\bm s} \mathbb{B}_{n}^{\bm s} \prod_k \Big[\frac{(1+s_k C_k)}{2}\Big]\nonumber\\
            = &\sum_{i\cdots jk} \alpha_{i\cdots jk} A_i \cdots B_j \Big[\sum_{\bm s} s_k \prod_{k'} \Big( \dfrac{1+s_{k'} C_{k'}}{2}\Big) \Big],
        \end{align}
        which leads to $\mathbb{B}_{n+1}$ by inserting Eq.~\eqref{eq:expansion}. 
    \end{proof}
    
    Since $\mathbb{B}_{n}^{\bm s}$ is obtained by assigning a specific value $s_k$ to each $C_k$ in $\mathbb{B}_{n+1}$, the maximal LHV value of $\mathbb{B}_{n}^{\bm s}$  cannot exceed the one of $\mathbb{B}_{n+1}$. That is, $\mathbb{B}_{n+1} {\leq l}$ implies $\mathbb{B}_{n}^{\bm s} {\leq l}$ for any LHV model. However, the exact LHV bound of $\mathbb{B}_{n}^{\bm s}$ for a given ${\bm s}$ might be strictly smaller than $l$.  

    Notice that the $\mathbb{B}_{n}^{\bm s}$'s are not independent of each other. They satisfy extra conditions such that the higher-order terms like $C_k C_{k'}$ are eliminated. Conversely, this inspires a general formula to build $(n+1)$-partite Bell inequalities from the $n$-partite ones.
    \begin{theorem}~\label{ob:iterative}
        For a given set of $n$-partite normalized Bell inequalities $\{\mathbb{B}_{n}^{\bm s} \leq 1\}_{{\bm s}}$, a $(n+1)$-partite normalized Bell inequality can be constructed as
        \begin{equation}\label{eq:IteratForm}
            \mathbb{B}_{n+1} = \frac{1}{2^m}\sum_{\bm s} \mathbb{B}_{n}^{\bm s} \Big(1+\sum_k s_k C_k\Big)\leq 1,
        \end{equation}
        if for any subset $K$ of $\{1, \cdots, m\}$, which contains at least two elements, we have
        \begin{equation}\label{eq:linearconditions}
            \sum\nolimits_{\bm s} \mathbb{B}_{n}^{\bm s} \Big(\prod\nolimits_{k\in K} s_k\Big) = 0.
        \end{equation}
    \end{theorem}
    \begin{proof}
        The conditions in Eq.~\eqref{eq:linearconditions} imply that
        \begin{align}
            \sum_{\bm s} \mathbb{B}_{n}^{\bm s} \prod_k (1+s_k C_k) &= \sum_{{\bm s},K} \mathbb{B}_{n}^{\bm s} \prod_{k\in K} (s_k C_k)\nonumber\\
            &=\sum_{{\bm s}, |K|\le 1} \mathbb{B}_{n}^{{\bm s}} \prod_{k\in K} (s_k C_k),
        \end{align}
       where $\prod_{k\in K} (s_k C_k) =1$ if $K=\emptyset$. Consequently, 
       $\mathbb{B}_{n+1} = \sum_{\bm s} \mathbb{B}_{n}^{\bm s} \prod_k \left[(1+s_k C_k)/{2}\right]$. From this and the condition
       in Eq.~(\ref{eq:linearconditions}) one can directly calculate that
       if one starts with $\mathbb{B}_{n+1}$ and computes the expression 
       $\tilde{\mathbb{B}}_{n}^{\bm s} \equiv\sum_{i\cdots jk} \alpha_{i\cdots jk} \,s_k A_i \cdots B_j$ as in Observation~\ref{observation1}, one finds that $\tilde{\mathbb{B}}_{n}^{\bm s}  = {\mathbb{B}}_{n}^{\bm s}.$
       Consequently, as the maximal LHV value 
       {of any of $\mathbb{B}_{n}^{\bm s}$ is $1$ and they are obtained from $\mathbb{B}_{n+1}$
       by fixing the values of the $C_k$}, it follows that the one of $\mathbb{B}_{n+1}$ is also $1$. 
       That is, $\mathbb{B}_{n+1}$ is also normalized. 
       
        Similarly, the maximal quantum value of $\mathbb{B}_{n+1}$ is no less than the one of any $\mathbb{B}_{n}^{\bm s}$, since we can always recover the quantum operator $\mathbb{B}_{n}^{\bm s}$ by setting $C_k = s_k \openone$, where $\openone$ signifies identity operator.
    \end{proof}
    Note that $\mathbb{B}_{n+1}$ constructed in Observation~\ref{ob:iterative} is nontrivial in 
    the sense that its quantum value is strictly larger than $1$, once at least one of $\mathbb{B}_n$ are nontrivial in the same sense. 
    This fails for the property of tightness (the property of being facets on the local polytope), namely if all $\mathbb{B}_n$ are tight, the $\mathbb{B}_{n+1}$ is not necessarily tight (a counterexample 
    is listed in Sec.~I of Supplemental Material (SM)~\cite{SM}).

    The inequality in Eq.~\eqref{eq:IteratForm} is the generalized iterative formula under the constraints in Eq.~\eqref{eq:linearconditions}. In the $(n+1,2,2)$ scenario, the only candidate of $K$ is $\{1,2\}$, then the resulting extra constraint is $\mathbb{B}_{n}^{(++)} +\mathbb{B}_{n}^{(--)} =\mathbb{B}_{n}^{(+-)} +\mathbb{B}_{n}^{(-+)}$.
    Consequently, there are still three $\mathbb{B}_{n}^{(\pm\pm)}$'s independent of each other, and under this circumstance,
     \begin{align}\label{eq:NAd122}
        \mathbb{B}_{n+1} =\dfrac{1}{2} \Bigl[& \mathbb{B}_n^{(++)} (C_1+C_2) +\mathbb{B}_n^{(+-)} (1-C_2) +\notag \\
            & \mathbb{B}_n^{(-+)} (1-C_1)\Bigl]\le 1.
     \end{align}
    We remark that the iterative formula in Observation~\ref{ob:iterative} cannot recover all the $(n+1)$-partite Bell inequalities, since we have only employed normalized $\mathbb{B}_{n}^{\bm s}$'s. However, by imposing other linear or symmetric constraints, we can recover well-known iterative Bell inequalities in the literature \cite{1990PRLM,1992PRAMA,1993PUBK,2006CAF,2003Sliwa,1981Froissart,2005PRLBIGS,2007WBZ}.

    \emph{Recovering Previous Iterative Formulas and beyond}.---One special solution to the constraint in Eq.~\eqref{eq:NAd122} for the $(n+1,2,2)$ scenario is that $\mathbb{B}_n^{(--)} = -\mathbb{B}_n^{(++)}$ and $\mathbb{B}_n^{(-+)} = -\mathbb{B}_n^{(+-)}$. In this case, the iterative formula in Observation~\ref{ob:iterative} becomes
    \begin{equation}\label{eq:general}
        \mathbb{B}_{n+1} \!=\! \frac{1}{2}\left[\mathbb{B}_n^{(++)}(C_1\!+\!C_2)\! +\! \mathbb{B}_n^{(+-)}(C_1\!-\!C_2)\right] \!\le\! 1.
    \end{equation}
    Starting with $\mathbb{B}_2^{(++)} \leq 1$ being the CHSH inequality 
    \begin{equation}\label{eq:CHSH}
        {\rm CHSH}=\dfrac{1}{2}\bigl(A_1 B_1+A_1 B_2+A_2 B_1-A_2 B_2\bigr)\leq 1,
    \end{equation}
    the iterative formula of the MABK inequalities \cite{1990PRLM,1992PRAMA,1993PUBK} can be recovered by taking $\mathbb{B}_n^{(++)}$ to be the MABK inequality for the $(n,2,2)$ scenario, and $\mathbb{B}_n^{(+-)}$ to be obtained from $\mathbb{B}_n^{(++)}$ by swapping the two measurements for each party, e.g., $A_1 \leftrightarrow A_2$.
    Similarly, if we take $\mathbb{B}_{n}^{(++)} \leq 1$ to be the $n$-partite MABK inequality and $\mathbb{B}_{n}^{(+-)} = \openone$, we obtain the CAF iterative formula~\cite{2006CAF}.

\begin{figure}[t]
          \centering
          \includegraphics[width = 1\columnwidth]{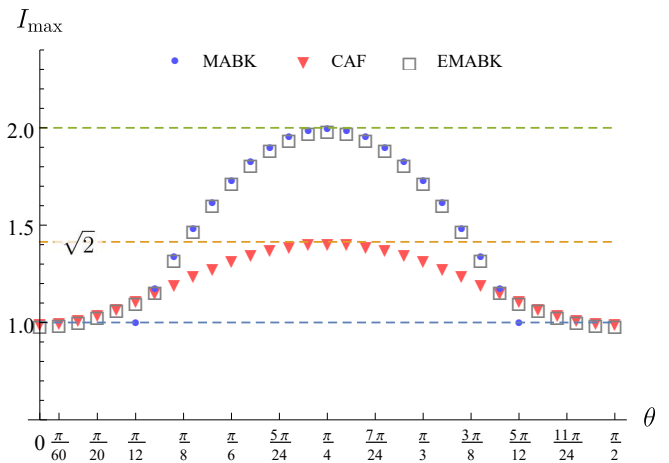}
          \caption{Maximal violations of the tripartite MABK, CAF and EMABK inequalities by using the state
             $\cos(\theta)\ket{000}+\sin(\theta)\ket{111}$ with $\theta\in[0,\pi/2]$. Note that the MABK inequality (blue dot) shows the maximal quantum violations $2$, but it is violated only in the region $\theta\in(\pi/12, 5\pi/12)$; the CAF inequality (red triangle) is violated in the whole entangled region, however, its maximal quantum violation is only $\sqrt{2}$; In comparison, the tripartite EMABK inequality (gray square) is a ``dual-use'' inequality, in the sense that it has the advantages of the MABK and CAF inequalities simultaneously.}
             \label{fig:CAF-MABK}
       \end{figure}

    Moreover, the generalized GHZ state $|\Phi(\theta)\rangle = \cos(\theta)\ket{000}+\sin(\theta)\ket{111}$ is a good testbed to observe the quantum behaviors with a given Bell inequality. The $n$-partite MABK inequality reaches its maximal quantum violation $2^{(n-1)/2}$ \cite{2001PRAWernerWolf} when $\theta=\pi/4$. However, for CAF inequalities, the maximal violation is only $2^{(n-2)/2}$~\cite{2006CAF}. The advantage of CAF inequalities is that they are always violated for the whole entangled region $\theta\in (0,\pi/2)$. In comparison, there exists some ranges of $\theta$ without quantum violation of the MABK inequalities, see Fig. \ref{fig:CAF-MABK} for more details of the tripartite case. As we have seen, the MABK inequalities and the CAF inequalities have their own advantages, which are not shared by each other. This intrigues one important question: Can one find out some ``dual-use'' Bell inequalities, in the sense that the following two properties are satisfied simultaneously: (i) Their maximal quantum violations are not weaker than the ones of MABK inequalities; (ii) They have the quantum violation in the whole entangled range of $\theta$? 

\emph{``Dual-Use'' Bell Inequalities.}---
    Bell inequalities with more measurements can reveal more entanglement usually. We consider the iterative formula in Eq.~\eqref{eq:general} in the $(3,3,2)$ scenario, which leads to one ``dual-use'' inequality. More explicitly, we take $\mathbb{B}_{n}^{(++)} \leq 1$ to be the standard CHSH inequality in Eq.~\eqref{eq:CHSH}, and $\mathbb{B}_{n}^{(+-)}$ is obtained from $\mathbb{B}_{n}^{(++)}$ by changing $A_1$ to $A_3$ and $A_2$ to $A_t$ ($t=1$ if $n$ is odd, and $t=4$ otherwise), same for other parties. For convenience, we name this inequality the extended-MABK (EMABK) inequality. 
       
    Furthermore, the tripartite EMABK inequality can be generalized into one iterative formula where the $(n+1)$-partite inequality is always of ``dual use''.
    \begin{theorem}\label{ob:emabk}
        Let $\mathbb{B}_{n}^{(++)} \leq 1$ be the standard $n$-partite MABK inequality, and $\mathbb{B}_{n}^{(+-)}$ be obtained from $\mathbb{B}_{n}^{(++)}$ by $M_1^{(i)}\to M_3^{(i)}$ and $M_2^{(i)}\to M_t^{(i)}$, where $M_k^{(i)}$ is the $k$-th measurement for the $i$-th party, $t=1$ if $n$ is odd, and $t=4$ otherwise. Then the $(n+1)$-partite Bell inequality defined in Eq.~\eqref{eq:general} has the ``dual-use'' property.
    \end{theorem}
    The proof of Observation~\ref{ob:emabk} is provided in Sec. II of SM \cite{SM}. We have one remark. The tripartite EMABK inequality is not the unique ``dual-use'' inequality if we allow more measurements for the third party, for example, the 
    Wie\'{s}niak-Badziag-\.{Z}ukowski (WBZ) inequality \cite{2007WBZ}. 

\emph{Three Measurements.---}
    Firstly, we can simplify the iterative formula in Observation~\ref{ob:iterative} with the linear constraints in Eq.~\eqref{eq:linearconditions}, which result in the following solution:
    \begin{align}\label{eq:3mSolution}
    \mathbb{B}_{n}^{(+--)} &=-\mathbb{B}_{n}^{(+++)} +\mathbb{B}_{n}^{(++-)} +\mathbb{B}_{n}^{(+-+)},\nonumber\\
    \mathbb{B}_{n}^{(-++)} &=2\,\mathbb{B}_{n}^{(+++)} -\mathbb{B}_{n}^{(++-)} -\mathbb{B}_{n}^{(+-+)} +\mathbb{B}_{n}^{(---)},\nonumber\\ 
    \mathbb{B}_{n}^{(-+-)} &=\mathbb{B}_{n}^{(+++)} -\mathbb{B}_{n}^{(+-+)} +\mathbb{B}_{n}^{(---)},\nonumber\\ 
    \mathbb{B}_{n}^{(--+)}&=\mathbb{B}_{n}^{(+++)} -\mathbb{B}_{n}^{(++-)} +\mathbb{B}_{n}^{(---)}.
    \end{align}
    By inserting this solution into Eq.~\eqref{eq:IteratForm}, we have the simplified general iterative formula, that is, 
    \begin{align}\label{eq:N32}
         \mathbb{B}_{n+1} =\dfrac{1}{2} \Bigl[& \mathbb{B}_{n}^{(+++)} (1-C_1+C_2+C_3) + \nonumber\\
            & \mathbb{B}_{n}^{(++-)} (C_1-C_3) +\mathbb{B}_{n}^{(+-+)} (C_1-C_2) + \nonumber\\
            & \mathbb{B}_{n}^{(---)} (1-C_1)\Bigl]\ \leq 1.
    \end{align}
    Notice that, even if all $\mathbb{B}_{n}^{(\pm\pm\pm)}$'s contain only $n$-partite correlations terms, there could still be terms in $\mathbb{B}_{n+1}$ which is not $n$-partite correlations. This issue can be overcome by introducing one extra condition that $\mathbb{B}_{n}^{(---)} =-\mathbb{B}_{n}^{(+++)}$. Then \eq{eq:N32} reduces to a simpler iterative formula as follows,
    \begin{align}\label{eq:N32FulCor}
        \mathbb{B}_{n+1} = \dfrac{1}{2} \Bigl[&\mathbb{B}_{n}^{(+++)} (C_2+C_3) +\mathbb{B}_{n}^{(++-)} (C_1-C_3) +\nonumber\\
            &\mathbb{B}_{n}^{(+-+)} (C_1-C_2)\Bigl]\ \leq 1.
    \end{align}
    For example, as a tripartite inequality with $n$-partite correlations terms, the WBZ inequality is recovered with $\mathbb{B}_{2}^{(+++)}$ being the standard CHSH inequality in Eq.~\eqref{eq:CHSH}, $\mathbb{B}_{2}^{(++-)}$ obtained from $\mathbb{B}_{2}^{(+++)}$ by $B_1 \to B_3$ and then $A \leftrightarrow B$, $\mathbb{B}_{2}^{(++-)}$ by $A_2\to A_3$ and $B_1\to B_3$.

    All the previous examples discussed in this work are based on the CHSH inequality, which contains only two measurements per party. Whereas the CHSH inequality is not the only relevant inequality in the $(2,3,2)$-scenario. The additional tight and relevant Bell inequality in the $(2,3,2)$-scenario is the I3322 inequality \cite{1981Froissart,2003Sliwa}, which can probe the nonlocality of some two-qubit states out of the capability of the CHSH inequality \cite{2004Collins}. Another interesting point of I3322 inequality is that its maximal quantum violation increases together with the dimension of the tested quantum system \cite{2010Pal}, which can identify the dimension of quantum system device-independently. Here we develop an iterative formula based on the I3322 inequality. For convenience, denote $\mathbb{B}^{\rm 3322}_n$ the $n$-partite I3322-Bell function according to this iterative formula, where $\mathbb{B}^{\rm 3322}_2 \leq 1$ is the standard normalized I3322 inequality \cite{1981Froissart}:
    \begin{align}
            \mathbb{B}_2^{\rm 3322} = \dfrac{1}{4} \Big[&A_1 -A_2 +B_1 -B_2 -(A_1-A_2)(B_1-B_2)  + \notag\\
                &(A_1 +A_2)B_3 +A_3(B_1 +B_2)\Big]\leq 1.
    \end{align}
    \begin{theorem}
    Let $\mathbb{B}_{n}^{(+++)}$ be the standard $\mathbb{B}^{\rm 3322}_n$ in the iterative formula, from which we obtain $\mathbb{B}_{n}^{(++-)}$ by $B_3 \rightarrow-B_3$, 
    $\mathbb{B}_{n}^{(+-+)}$ by $A_3 \rightarrow-A_3$, and 
    $\mathbb{B}_{n}^{(---)}$ by $A_1 \rightarrow-A_1, A_2 \rightarrow-A_2$ and $A_3 \rightarrow-A_3$, then we obtain the $(n+1)$-partite Bell inequality $\mathbb{B}^{\rm 3322}_{n+1}$ as in Eq. \eqref{eq:N32}. 
    \end{theorem}
    The explicit expression of the tripartite Bell inequality in this iteration reads
    \begin{align}\label{eq:i33223}
        \mathbb{B}_3^{\rm 3322}\! =\! \dfrac{1}{4} \Bigl[&(A_1\! -\!A_2)C_1\! +\!B_1\! -\!B_2\! -\!(A_1\!-\!A_2)(B_1\!-\!B_2)C_1\! + \notag\\
                &(A_1\!+\!A_2)B_3 C_3\! +\!A_3(B_1\! +\!B_2)C_2\Bigr]\leq 1.
    \end{align}
    By replacing $C_i$ with $C_iD_i$, it results in $\mathbb{B}^{\rm 3322}_{4}$. In the same way, we obtain $\mathbb{B}^{\rm 3322}_{n}$ for $n\ge 5$. As it turns out, $\mathbb{B}^{\rm 3322}_{n} \leq 1$ is still tight. Note that $\mathbb{B}^{\rm 3322}_n$ is not invariant under the permutation of parties any more, which is different from the generalizations of I3322 in Refs.~\cite{bancal2010looking,bernards2020generalizing,2021Bernards}.

\emph{Extending \'Sliwa's Inequalities.}---
    As we have seen in the previous sections, it is already powerful enough to use one $\mathbb{B}_n \leq 1$ and its equivalent Bell inequalities as generators for  $(n+1)$-partite Bell inequalities. For a given normalized $\mathbb{B}_n \leq 1$, the general procedure is as follows. Firstly, we generate the equivalence class of $\mathbb{B}_n$ by permutations of parties, measurements and outcomes. Then we choose all $\mathbb{B}_n^{\bm s}$ from this class and test the conditions in Eq.~\eqref{eq:linearconditions}. If all the conditions are fulfilled, we construct $(n+1)$-partite Bell inequality as in Eq.~\eqref{eq:IteratForm}. 

    In SM \cite{SM} (see Sec. III B), we have employed \'Sliwa's inequalities in such a way to generate four-partite tight ones. As an example, we present the first inequality in \'Sliwa's family, i.e., 
    ${\text{\'Sliwa}}_1 = A_1+B_1-A_1 B_1+C_1-A_1 C_1-B_1 C_1+A_1 B_1 C_1 \leq 1$, which is trivial from the perspective of quantum violation. Notwithstanding, let $\mathbb{B}_3^{(++)}$ be the standard $\text{\'Sliwa}_1$ in the iterative formula, from which we obtain $\mathbb{B}_{n}^{(+-)}$ by $C_1 \leftrightarrow C_2$, $\mathbb{B}_{n}^{(-+)}$ by $C_1 \leftrightarrow C_2$, then 
    $C_2 \rightarrow-C_2$, and $\mathbb{B}_{n}^{(--)}$ by $C_1 \rightarrow-C_1$, then we obtain the four-partite generalization of $\text{\'Sliwa}_1$ as in Observation~\ref{ob:iterative} with the measurements $D_1, D_2$ for the last party. More explicitly, the four-partite inequality reads
    \begin{align}
        \mathbb{B}_4 =\dfrac{1}{2}\Bigl[&\mathbb{B}_3^{(++)} (D_1+D_2) +\mathbb{B}_3^{(+-)} (1-D_2) +\nonumber\\
            &\mathbb{B}_3^{(-+)} (1-D_1)\Bigl]\ \leq 1,
    \end{align}
    whose maximal quantum violation is $4\sqrt{2}-3$.

\emph{Conclusion and Discussion.}---
    In this work, we have developed a very general iterative formula to generate multipartite Bell inequalities from few-partite ones. By imposing extra linear conditions and symmetries on the $n$-partite ones, the iterative formula can also be simplified. Based on this observation and starting from the CHSH inequality, we have not only recovered famous families of multipartite Bell inequalities, like the MABK inequalities and the CAF inequalities, but also found the new EMABK inequalities to combine the advantages of the aforementioned two families. Starting with I3322 and the inequalities discovered by \'Sliwa, new inequalities with the properties like tightness have also been constructed. Hence our general iterative formula is powerful in the field of Bell nonlocality.

    We have concerned ourselves mainly with the Bell inequalities in the $(n,m,2)$-scenarios here, while the generalization to the $(n,m,d)$-scenarios or other concepts like network nonlocality is yet to be done. Besides, an effective algorithm to implement the iterative formula starting from a given Bell inequality and its equivalence is also desired. Finally, as Bell inequalities are closely connected to information processing tasks, it would be very interesting to clarify in which sense our iteration method allows to identify novel quantum resources.

\emph{Acknowledgements}.---
    J. L. C. was supported by the National Natural Science Foundations of China ( Grant Nos. 12275136, 11875167 and 12075001), the 111 Project of B23045, and the Fundamental Research Funds for the Central Universities (Grant No. 3072022TS2503). O.G. and Z.P.X. acknowledge the support from the Deutsche Forschungsgemeinschaft (DFG, German Research Foundation, project numbers 447948357 and 440958198), the Sino-German Center for Research Promotion (Project M-0294), the ERC (Consolidator Grant 683107/TempoQ), the German Ministry of Education and Research (Project QuKuK, BMBF Grant No.~16KIS1618K), and the Alexander Humboldt foundation. X.Y.F was supported by the Nankai Zhide Foundation. H.X.M. was supported by the National Natural Science Foundations of China (Grant No. 11901317). 
\bibliography{main}
\end{document}


\title{Supplemental Material for ``Generalized Iterative Formula for Bell Inequalities''}

\author{Xing-Yan Fan}
\affiliation{Theoretical Physics Division, Chern Institute of Mathematics, Nankai 
  University, Tianjin 300071, People's Republic of China}

\author{Zhen-Peng Xu}
\email{zhen-peng.xu@ahu.edu.cn}
\affiliation{School of Physics and Optoelectronics Engineering, Anhui University, 230601 Hefei, People's Republic of China}

\author{Jia-Le Miao}
\affiliation{CAS Key Laboratory of Quantum Information, University of Science and Technology of China, Hefei, 230026, People's Republic of China}

\author{Hong-Ye Liu}
\affiliation{CAS Key Laboratory of Theoretical Physics, Institute of Theoretical Physics, Chinese Academy of Sciences, Beijing 100190, China}

\author{Yi-Jia Liu}
\affiliation{CAS Key Laboratory of Quantum Information, University of Science and Technology of China, Hefei, 230026, People's Republic of China}

\author{Wei-Min Shang}
\affiliation{School of Science, Tianjin Chengjian University, Tianjin 300384, People's Republic of China}
\affiliation{Theoretical Physics Division, Chern Institute of Mathematics, Nankai University, Tianjin 300071, People's Republic
  of China}

\author{Jie Zhou}
\affiliation{Theoretical Physics Division, Chern Institute of Mathematics, Nankai University, Tianjin 300071, People's Republic
  of China}

\author{Hui-Xian Meng}
\affiliation{School of Mathematics and Physics, North China Electric Power University, Beijing 102206, People's Republic of
  China}

\author{Otfried G\"{u}hne}
\email{otfried.guehne@uni-siegen.de}
\affiliation{Naturwissenschaftlich-Technische Fakult\"at, Universit\"at Siegen, 
  Walter-Flex-Stra\ss e 3, 57068 Siegen,Germany}

\author{Jing-Ling~Chen}
\email{chenjl@nankai.edu.cn}
\affiliation{Theoretical Physics Division, Chern Institute of Mathematics, Nankai 
  University, Tianjin 300071, People's Republic of China}

\date{\today}

\maketitle
\tableofcontents   
\section{An Counterexample of Non-Tight $\mathbb{B}_4$ from Tight $\mathbb{B}_3$}
    Let $\mathbb{B}_3^{(++)}$ be the \'Sliwa's first inequality \cite{2003Sliwa}, i.e.
    \begin{equation}
        \mathbb{B}_3^{(++)} =A_1 +B_1 +C_1 -A_1 C_1 -B_1 C_1 -A_1 B_1 +A_1 B_1 C_1\le 1,
    \end{equation}
    and through the symmetric transformations
    \begin{equation}\label{eq:SplitSliwa1B4}
        \mathbb{B}_3^{(+-)} =\mathbb{B}_3^{(++)} ,\quad \mathbb{B}_3^{(--)} =\mathbb{B}_3^{(-+)} =\mathbb{B}_3^{(++)} (C_1 \rightarrow-C_1),
    \end{equation} 
    then the iterative formula
    \begin{equation}
        \mathbb{B}_4 =\dfrac{1}{4} \Bigl[\mathbb{B}_3^{(++)} (1+D_1)(1+D_2)+\mathbb{B}_3^{(+-)} (1+D_1)(1-D_2)+\mathbb{B}_3^{(-+)} (1-D_1)(1+D_2)
            +\mathbb{B}_3^{(--)} (1-D_1)(1-D_2)\Bigl]\le 1
    \end{equation}
    tells us 
    \begin{equation}\label{eq:Sliwa1B4}
        \mathbb{B}_4 =A_1 +B_1 +C_1 D_1 -A_1 C_1 D_1 -B_1 C_1 D_1 -A_1 B_1 +A_1 B_1 C_1 D_1 \le 1.
    \end{equation}
    The inequality \eqref{eq:Sliwa1B4} is not tight from numerical check, while its split forms in \EquRef{eq:SplitSliwa1B4} are all tight.
\section{The Proof of Observation 3}
      Given a standard Clause-Horne-Shimony-Holt (CHSH) inequality \cite{1969CHSH} $(A_1 B_1+A_1 B_2+A_2 B_1-A_2 B_2)/2\le 1$, a complete group of its equivalent partners can be generated under the three kinds of transformations (permutations of parties, measurements or outcomes), that is
     \begin{align}\label{eq:SCHSH}
        \mathcal{S}_{\rm CHSH}:=\biggl\{&\dfrac{1}{2}\bigl(A_1 B_1-A_1 B_2-A_2 B_1-A_2 B_2\bigr)\le 1,\,
           \dfrac{1}{2}\bigl(-A_1 B_1-A_1 B_2+A_2 B_1-A_2 B_2\bigr)\le 1, \notag\\ 
           &\dfrac{1}{2}\bigl(-A_1 B_1+A_1 B_2-A_2 B_1-A_2 B_2\bigr)\le 1,\,
           \dfrac{1}{2}\bigl(A_1 B_1+A_1 B_2+A_2 B_1-A_2 B_2\bigr)\le 1, \notag\\
           &\dfrac{1}{2}\bigl(-A_1 B_1-A_1 B_2-A_2 B_1+A_2 B_2\bigr)\le 1,\,
           \dfrac{1}{2}\bigl(A_1 B_1-A_1 B_2+A_2 B_1+A_2 B_2\bigr)\le 1, \notag\\
           &\dfrac{1}{2}\bigl(A_1 B_1+A_1 B_2-A_2 B_1+A_2 B_2\bigr)\le 1,\,
           \dfrac{1}{2}\bigl(-A_1 B_1+A_1 B_2+A_2 B_1+A_2 B_2\bigr)\le 1\biggr\}.
     \end{align}
     For convenience, let $\mathcal{S}_{\rm CHSH} [i]$ denote the $i$-th entity in the set $\mathcal{S}_{\rm CHSH}$, $i\in\set{1,2,...,8}$, thus $\mathcal{S}_{\rm CHSH} [4]$ refers to the standard CHSH inequality. Arbitrary $(n+1)$-partite Mermin-Ardehali-Belinski{\u{\i}}-Klyshko (MABK) inequalities \cite{1990PRLM,1992PRAMA,1993PUBK} are constructed as follows. (Commonly) choose $\mathbb{B}_2^{(++)} ={\rm CHSH}$ as the standard one in \eqref{eq:SCHSH} (i.e., ${\rm CHSH}=\mathcal{S}_{\rm CHSH} [4]$) and $\mathbb{B}_2^{(+-)}$ to be obtained from $\mathbb{B}_2^{(++)}$ by swapping the two measurements for parties $A$ and $B$, i.e., $A_1 \leftrightarrow A_2$ and $B_1 \leftrightarrow B_2$. Through 
    \begin{equation}
        \mathbb{B}_3 =\dfrac{1}{2} \Bigl[\mathbb{B}_2^{(++)} (C_1+C_2)+\mathbb{B}_2^{(+-)} (C_1-C_2)\Bigl]\le 1,
     \end{equation}
      we get the tripartite MABK inequality. Similarly, let $\mathbb{B}_n^{(++)}$ be the $n$-partite MABK polynomial, $\mathbb{B}_n^{(+-)}$ to be obtained from 
      $\mathbb{B}_n^{(++)}$ by swapping the two measurements for each party. Finally using \EquRef{eq:general} completes the iteration.

      The \textbf{Observation 3} in main text reads:

      \emph{Let $\mathbb{B}_{n}^{(++)} \le 1$ be the standard $n$-partite MABK 
         inequality, and $\mathbb{B}_{n}^{(+-)}$ be obtained from 
         $\mathbb{B}_{n}^{(++)}$ by $M_1^{(i)}\to M_3^{(i)}$ and 
         $M_2^{(i)}\to M_t^{(i)}$ where $M_k^{(i)}$ is the $k$-th measurement for the $i$-th party, $t=1$ if $n$ is odd, and $t=4$ otherwise, then the $(n+1)$-partite extended MABK (EMABK) inequality defined via
         \begin{equation}\label{eq:general}
            \mathbb{B}_{n+1} =\dfrac{1}{2} \left[\mathbb{B}_n^{(++)} (C_1+C_2) 
               +\mathbb{B}_n^{(+-)} (C_1-C_2)\right]\le 1
         \end{equation}
         is ``dual-use''.}
      \subsection{Predefinitions and Analyses}
         Quantum mechanically, for the observables $A_l$, $B_l$, and $C_l$ with two measurement outcomes, 
         \begin{align*}
            A_l \equiv\vec{\sigma}\cdot(\sin\theta_{al}\cos\varphi_{al},
               \sin\theta_{al}\sin\varphi_{al},\cos\theta_{al})
            =\begin{bmatrix}
               \cos\theta_{al} & \sin\theta_{al}\:{\rm e}^{-{\rm i}\,\varphi_{al}} \\
               \sin\theta_{al}\:{\rm e}^{{\rm i}\,\varphi_{al}} & -\cos\theta_{al}
            \end{bmatrix},
         \end{align*}
         so do $B_l$ and $C_l$, $l\in\set{1,2,3}$. Without loss of generality, we fix all Bloch vectors relating to the observations shown in a typical Bell inequality on $xz$-plane, which means that $\varphi_{\rm m} =0$, here ``m'' expresses the parameter $\varphi_{\rm m}$ involving ``measurements'',  i.e.,
         \begin{equation}\label{eq:AlPhi0}
            A_l =\begin{bmatrix}
               \cos\theta_{al} & \sin\theta_{al} \\
               \sin\theta_{al} & -\cos\theta_{al}
            \end{bmatrix},
         \end{equation}
         so do $B_l$ and $C_l$, $l\in\set{1,2,3}$. Of course, there is another way to simplify calculation, namely setting $\theta_{\rm m} =\pi/2$ ($xy$-plane), then 
         \begin{equation}\label{eq:AlThetPiO2}
            A_l =\begin{bmatrix}
               0 & {\rm e}^{-{\rm i}\,\varphi_{al}} \\
               {\rm e}^{{\rm i}\,\varphi_{al}} & 0
            \end{bmatrix},
         \end{equation}
         so do $B_l$ and $C_l$, $l\in\set{1,2,3}$. 
         
         To accomplish the demonstration, we need to prove the following two lemmas for even and odd $n$ respectively.
         \begin{lemma}\label{lem:1}
            The maximal quantum violation of EMABK inequality $\mathbb{B}_n \le 1$ is as strong as that of the $n$-partite MABK inequality, i.e., 
            $\mathbb{B}_n^{\rm max} =2^{(n-1)/2}$.
         \end{lemma}
         \begin{lemma}\label{lem:2}
            The $n$-partite EMABK inequality $\mathbb{B}_n \le 1$ is violated in the whole entangled region $\theta\in(0,\pi/2)$.
         \end{lemma}

         We have known that the maximal quantum violation of the (normalized) $n$-qubit MABK inequality is $2^{(n-1)/2}$ \cite{2001PRAWernerWolf}, iff the system is at the $n$-qubit Greenberger-Horne-Zeilinger (GHZ) state \cite{1989GHZ} or its unitary equivalent partners \cite{2004PRLZqC}. First we try to expain this fact using the generalized GHZ state 
         $\ket{\Psi_{\rm GGHZ}(\theta)}=\cos(\theta) \ket{00\cdots 0} +\sin(\theta) |11\cdots 1\rangle$, $\theta\in(0,\pi/2)$, (when $\theta=\pi/4$, that is the $n$-qubit GHZ state).

         For simplicity, we denote the antidiagonal elements $e_j$ of a given square matrix in column $j$ as 
         ${\rm Adiag}(e_1 ,e_2,...,e_{2^n -1} ,e_{2^n})$. Likewise, define ${\rm Diag}(e_1,e_2,...,e_{2^n -1},e_{2^n})$ as the diagonal elements of a matrix. Since we care about the elements posed on four corners in a specific matrix merely, the corner or (anti)diagonal elements can be used to mark a matrix thereafter.

         Select $\theta_{\rm m} =\pi/2$, and
         \begin{align*}
            & \varphi_{a1} =0,\ \varphi_{a2} =\dfrac{\pi}{2}, \\
            & \varphi_{b1} =\varphi_{c1} =\cdots=\varphi_{d1} =-\dfrac{\pi}{4}, \\
            & \varphi_{b2} =\varphi_{c2} =\cdots=\varphi_{d2} =\dfrac{\pi}{4},
         \end{align*}
         which indicates
         \begin{equation}
            \begin{split}
               & A_1 =\begin{bmatrix} 0 & 1 \\ 1 & 0 \end{bmatrix}={\rm Adiag}(1,1),\ 
                  A_2=\begin{bmatrix} 0 & -{\rm i} \\ {\rm i} & 0 \end{bmatrix}
                     ={\rm i}\,{\rm Adiag}(1,-1),\\
               & B_1 =C_1 =\cdots=D_1 =\dfrac{1}{\sqrt{2}} \begin{bmatrix}
                     0 & 1+\rm i \\
                     1-\rm i & 0 \end{bmatrix}=\dfrac{1}{\sqrt{2}} {\rm Adiag}(
                        1-{\rm i},1+{\rm i}),\\
               & B_2 =C_2 =\cdots=D_2 =\dfrac{1}{\sqrt{2}} \begin{bmatrix}
                     0 & 1-\rm i \\
                     1+\rm i & 0
                  \end{bmatrix}=\dfrac{1}{\sqrt{2}} {\rm Adiag}(1+{\rm i},1-{\rm i}). 
            \end{split}
         \end{equation}
         For example, when $n=2$,
         \begin{equation}\label{eq:MatCHSH}
            {\rm CHSH}\equiv\dfrac{1}{2} \Bigl[A_1 (B_1 +B_2) 
               +A_2 (B_1 -B_2)\Bigr]=\sqrt{2}\,{\rm Adiag}(1,0,0,1),
         \end{equation}
         so does 
         \begin{equation}
            {\rm CHSH}'\equiv\dfrac{1}{2} \Bigl[A_2 (B_1 +B_2)-A_1 (B_1 -B_2)\Bigr]
            =\sqrt{2}\,{\rm i}\,{\rm Adiag}(1,0,0,-1)
         \end{equation}
         up to a setting permutation cf. Eq. \eqref{eq:MatCHSH}. Recursively, we write down the tripartite MABK operator,
         \begin{equation}
            \mathbb{B}^{\rm MABK}_3 \equiv\dfrac{1}{2} \Bigl[{\rm CHSH} \bigl(C_1 +C_2\bigr)
               +{\rm CHSH}'\bigl(C_1 -C_2\bigr)\Bigr]
            =2\,{\rm Adiag}(1,0,0,0,0,0,0,1).
         \end{equation}
         Through the iteration above, the $n$-qubit $2^n$-dimensional MABK operator reads 
         \begin{equation}
            \mathbb{B}^{\rm MABK}_n \equiv 2^{(n-1)/2}\,{\rm Adiag}(1,0,...,0,1),
         \end{equation}
         which can be maximized to $2^{(n-1)/2}$ for the $n$-qubit GHZ state 
         $\bigl(|00\cdots 0\rangle+|11\cdots 1\rangle\bigr)/\sqrt{2}$.
      \subsection{Proof for Even $n$}
         \subsubsection{Lemma 1}
            \begin{proof} 
               For \LemRef{lem:1}, we find that under the assumption of 
               \eqref{eq:AlThetPiO2}, $\exv{\mathbb{B}_n}\le 2^{(n-1)/2}$, and if
               \begin{equation}\label{eq:PhiEveN}
                  \begin{split}
                     & \varphi_{a1} =0,\ \varphi_{a2} =\varphi_{a3} =\dfrac{\pi}{2}, \\
                     & \varphi_{b1} =\varphi_{c1} =\cdots=\varphi_{d1} =
                        -\dfrac{\pi}{4}, \\
                     & \varphi_{b2} =\varphi_{b3} =\varphi_{c2} =\varphi_{c3}
                        =\cdots=\varphi_{d2} =\varphi_{d3} =\dfrac{\pi}{4},
                  \end{split}
               \end{equation}
               the maximal violation is gained.
               
               Since we have known that in MABK inequality
               \begin{equation}\label{eq:MABK}
                  \begin{split}
                     \mathbb{B}^{\rm MABK}_n =\dfrac{1}{2} \Bigl[
                           \mathbb{B}^{\rm MABK}_{n-1} (D_1 +D_2)
                           +\bigl(\mathbb{B}^{\rm MABK}_{n-1}\bigr)'(D_1-D_2)\Bigr]
                        \le  1,
                  \end{split}
               \end{equation}
               and $\bigl(\mathbb{B}^{\rm MABK}_{n-1}\bigr)'$ is obtained from $\mathbb{B}^{\rm MABK}_{n-1}$ via a 
               permutation for measurement settings. Although 
               $\mathbb{B}_{n-1}^{(+-)} =\mathbb{B}_{n-1}^{(++)} (A_1\to A_3, 
                  A_2\to A_1, B_1\to B_3, B_2\to B_1,\cdots, 
                  D_1 \to D_3, D_2 \to D_1)$, the matrix form of the operators 
               $\bigl(\mathbb{B}^{\rm MABK}_{n-1}\bigr)'$ and $\mathbb{B}_{n-1}^{(+-)}$ in \textbf{Observation 3} are uniform under the constraint 
               \eqref{eq:PhiEveN}. Hence 
               $\exv{\mathbb{B}_n^{\rm EMABK}}=\exv{\mathbb{B}^{\rm MABK}_n}$ for the $n$-qubit GHZ state 
               $\bigl(|00\cdots 0\rangle+|11\cdots 1\rangle\bigr)/\sqrt{2}$. 
               This ends the proof of \LemRef{lem:1} for even $n$.
            \end{proof}
         \subsubsection{Lemma 2}
            As for \LemRef{lem:2}, we notice once condition \eqref{eq:AlPhi0} holds, together with 
            \begin{equation}
               \begin{split}
                  & \theta_{a1} =\theta_{a2} =\dfrac{\pi}{2},\ \theta_{a3}=0,\\
                  & \theta_{b2} =\cdots=\theta_{c2} =\dfrac{\pi}{2},\ 
                     \theta_{b1} =\cdots=\theta_{c1} 
                     =\theta_{b3} =\cdots=\theta_{c3} =0, \\
                  & \theta_{d2} =\pi-\theta_{d1},
               \end{split}
            \end{equation}
            namely
            \begin{equation}\label{eq:MaTheta}
               \begin{split}
                  & A_1 =A_2 =\begin{bmatrix} 0 & 1 \\ 1 & 0 \end{bmatrix},\ 
                     A_3 =\begin{bmatrix} 1 & 0 \\ 0 & -1 \end{bmatrix}, \\
                  & B_2 =\cdots=C_2 =\begin{bmatrix} 0 & 1 \\ 1 & 0 \end{bmatrix},\ 
                     B_1 =\cdots=C_1 =B_3 =\cdots=C_3 
                        =\begin{bmatrix} 1 & 0 \\ 0 & -1\end{bmatrix}, \\
                  & D_1 =\begin{bmatrix} \cos\theta_{d1} & \sin\theta_{d1} \\ 
                        \sin\theta_{d1} & -\cos\theta_{d1}\end{bmatrix},\ 
                     D_2 =\begin{bmatrix} -\cos\theta_{d1} & \sin\theta_{d1} \\ 
                        \sin\theta_{d1} & \cos\theta_{d1}\end{bmatrix},
               \end{split}
            \end{equation}
            then $\exv{\mathbb{B}_n}>1$ is satisfied in the whole entangled region $\theta\in(0,\pi/2)$.
            \begin{proof}
               When $n=2$, for $\theta_{a1}=\pi/2$, $\theta_{a3}=0$, and $\theta_{b2}=\pi-\theta_{b1}$, we obtain 
               \begin{equation*}
                  A_1=\begin{bmatrix}0 & 1 \\ 1 & 0\end{bmatrix},\ A_3=\begin{bmatrix}1 & 0 \\ 0 & -1\end{bmatrix},\ 
                  B_1=\begin{bmatrix}\cos\theta_{b1} & \sin\theta_{b1} \\ \sin\theta_{b1} & -\cos\theta_{b1}\end{bmatrix},\ 
                  B_2=\begin{bmatrix}-\cos\theta_{b1} & \sin\theta_{b1} \\ \sin\theta_{b1} & \cos\theta_{b1}\end{bmatrix},
               \end{equation*}
               which implies 
               \begin{equation}
                  {\rm CHSH}=\dfrac{1}{2} \Bigl[A_1 (B_1 +B_2)
                     +A_3(B_1-B_2)\Bigr]=\begin{bmatrix}
                        \cos\theta_{b1} & 0 & 0 & \sin\theta_{b1} \\
                        0 & -\cos\theta_{b1} & \sin\theta_{b1} & 0 \\
                        0 & \sin\theta_{b1} & -\cos\theta_{b1} & 0 \\
                        \sin\theta_{b1} & 0 & 0 & \cos\theta_{b1}
                     \end{bmatrix}.
               \end{equation}
               After that, the mean value of CHSH operator for the state 
               $\ket{\Psi(\theta)}=\cos\theta \ket{00}+\sin\theta \ket{11}$ reads 
               \begin{equation}
                  \exv{{\rm CHSH}} =\cos\theta_{b1} +\sin(2\theta)\sin\theta_{b1}
                  =\sqrt{1+\sin^2(2\theta)}\cos(\theta_{b1}-\eta),         
               \end{equation}
               where $\tan\eta=\sin(2\theta)$. In this way, set $\theta_{b1}=\eta$, then $\forall\,\theta\in(0,\pi/2)$, $\exv{\rm CHSH}>1$ all the time. Notice for every $n$ (e.g., $n=2$ for CHSH), the two measurement settings of the last party are completely same, therefore we may need to construct 
               $\mathbb{B}_{n-1}^{(++)}$ with the nonzero antidiagonal elements $k_1$ in both column 1 and $2^{n-1}$, similarly for the nonzero diagonal elements of $\mathbb{B}_{n-1}^{(+-)}$ in column 1 valued $k_2$ and column $2^{n-1}$ valued $-k_2$, i.e.,
               \begin{equation}\label{eq:B1B2}
                  \mathbb{B}_{n-1}^{(++)} =k_1\,{\rm Adiag}(1,...,1),\ 
                  \mathbb{B}_{n-1}^{(+-)} =k_2\,{\rm Diag}(1,...,-1),
               \end{equation}
               with $k_1 ,k_2$ implying two constants, inspired via the expressions of $A_1$ and $A_3$.

               In the case of $n=4$, select the following measurements
               \begin{align*}
                  \theta_{a1} =\theta_{a2} =\dfrac{\pi}{2},\ \theta_{a3}=0,\
                  \theta_{b2} =\theta_{c2} =\dfrac{\pi}{2},\ 
                  \theta_{b1} =\theta_{c1} =\theta_{b3} =\theta_{c3} =0,\ 
                  \theta_{d2} =\pi-\theta_{d1},
               \end{align*}
               then 
               \begin{align*}
                  &A_1=A_2=\begin{bmatrix}0 & 1 \\ 1 & 0\end{bmatrix},\ A_3=\begin{bmatrix}1 & 0 \\ 0 & -1\end{bmatrix},\\
                  & B_2 =C_2 =\begin{bmatrix}0 & 1 \\ 1 & 0\end{bmatrix},\ 
                     B_1 =B_3 =C_1 =C_3 =\begin{bmatrix}1 & 0 \\ 0 & -1\end{bmatrix},\\
                  &D_1=\begin{bmatrix}\cos\theta_{d1} & \sin\theta_{d1} \\ \sin\theta_{d1} & -\cos\theta_{d1}\end{bmatrix},\ 
                     D_2=\begin{bmatrix}-\cos\theta_{d1} & \sin\theta_{d1} \\ \sin\theta_{d1} & \cos\theta_{d1}\end{bmatrix}.
               \end{align*}
               Further we obtain 
               \begin{align*}
                  &\mathbb{B}_2^{(++)} ={\rm CHSH}=\dfrac{1}{2} \Bigl[A_1 (B_1+B_2)
                     +A_2 (B_1-B_2)\Bigr]=A_1 B_1
                  =\begin{bmatrix} 0 & 1 \\ 1 & 0 \end{bmatrix}\otimes\begin{bmatrix}
                     1 & 0 \\ 0 & -1 \end{bmatrix} \\
                  &\mathbb{B}_2^{(+-)}=\dfrac{1}{2}\Bigl[A_2(B_1+B_2)+A_1(B_2-B_1)\Bigr]=A_2 B_2
                  =\begin{bmatrix}0 & 1 \\ 1 & 0\end{bmatrix}\otimes\begin{bmatrix}0 & 1 \\ 1 & 0\end{bmatrix}={\rm Adiag}(1,1,1,1),
               \end{align*}
               and 
               \begin{align*}
                  \mathbb{B}_2^{(++)'}:= &\mathbb{B}_2^{(++)} \bigl(
                     A_{1\to 3} ,A_{2\to 1} ,B_{1\to 3} ,B_{2\to 1}\bigr) 
                  =\dfrac{1}{2} \Bigl[A_3(B_3+B_1)+A_2(B_3-B_1)\Bigr]=A_3 B_3
                  =\begin{bmatrix} 1 & 0 \\ 0 & -1\end{bmatrix}\otimes\begin{bmatrix}
                     1 & 0 \\ 0 & -1\end{bmatrix} \\
                  =& {\rm Diag}(1,-1,-1,1), \\
                  \mathbb{B}_2^{(+-)'} :=& \mathbb{B}_2^{(+-)} \bigl(
                        A_{1\to 3} ,A_{2\to 1} ,B_{1\to 3} ,B_{2\to 1}
                     =\dfrac{1}{2} \Bigl[A_2 (B_3+B_1)+A_3 (B_1-B_3)\Bigr]=A_2 B_3
                  =\begin{bmatrix} 0 & 1 \\ 1 & 0 \end{bmatrix}\otimes\begin{bmatrix}
                     1 & 0 \\ 0 & -1\end{bmatrix},\\
                  C_3 +C_1 =& 2\begin{bmatrix} 1 & 0 \\ 0 & -1 \end{bmatrix},\
                  \mathbb{B}_3^{(+-)} :=\dfrac{1}{2} \Bigl[
                     \mathbb{B}_2^{(++)'} (C_3+C_1)+\mathbb{B}_2^{(+-)'} (C_3-C_1)
                     \Bigr]=\mathbb{B}_2^{(++)'} \otimes\begin{bmatrix}
                        1 & 0 \\ 0 & -1\end{bmatrix}.
               \end{align*}
               After that, the matrix forms of $\mathbb{B}_3^{(++)}$, and 
               $\mathbb{B}_3^{(+-)}$ are as follows. 
               \begin{equation}\label{eq:N3B1}
                  \mathbb{B}_3^{(++)} :=\mathbb{B}^{\rm MABK}_3=\dfrac{1}{2}\Bigl[
                     \mathbb{B}_2^{(++)}(C_1+C_2)+\mathbb{B}_2^{(+-)}(C_1-C_2)\Bigr]
                  =-\dfrac{1}{2} \begin{bmatrix} 0 & M \\ M & 0 \end{bmatrix}
                  =-\dfrac{1}{2} \begin{bmatrix}
                     0 & \cdots & 1 \\
                     \vdots & \ddots & \vdots \\
                     1 & \cdots & 0  
                  \end{bmatrix}_{8\times 8},
               \end{equation}
               with 
               \begin{equation}
                  M\equiv\begin{bmatrix}
                     1 & 1 & 1 & -1 \\
                     1 & -1 & -1 & -1 \\
                     1 & -1 & -1 & -1 \\
                     -1 & -1 & -1 & 1
                  \end{bmatrix};
               \end{equation}
               and
               \begin{equation}\label{eq:N3B2}
                  \mathbb{B}_3^{(+-)} :=\mathbb{B}_3^{(++)} \bigl(A_{1\to 3},
                     A_{2\to 1} ,B_{1\to 3} ,B_{2\to 1} ,C_{1\to 3} ,C_{2\to 1}\bigr)
                  ={\rm Diag}(1,-1,-1,1,-1,1,1,-1).
               \end{equation}
               In view of the fact that we concern ourselves with the (anti)diagonal elements in \EqsRef{eq:N3B1}{eq:N3B2}, thus they match \EquRef{eq:B1B2} with $k_1=-1/2$ and $k_2=1$. Further, we discover
               \begin{equation*}
                  \mathbb{B}_{n-1}^{(++)} :=\mathbb{B}^{\rm MABK}_{n-1} =\dfrac{1}{2}\Bigl[\mathbb{B}_{n-2}^{(++)} \bigl(C_1+C_2\bigr)
                     +\mathbb{B}_{n-2}^{(+-)} \bigl(C_1-C_2\bigr)\Bigr],
               \end{equation*}
               and $C_1 -C_2 =\begin{bmatrix} 1 & -1 \\ -1 & -1 \end{bmatrix}$ holds for any even $n$. Hence according to the iterative formula of MABK inequality, 
               \begin{equation}
                  \mathbb{B}_{n-1}^{(++)}
                  =-\dfrac{1}{2^{(n-2)/2}}\begin{bmatrix}
                     0 & \cdots & 1 \\
                     \vdots & \ddots & \vdots \\
                     1 & \cdots & 0  
                  \end{bmatrix}_{2^{n-1}\times 2^{n-1}}.
               \end{equation}
               Likewise, 
               \begin{align*}
                  \mathbb{B}_{n-1}^{(+-)} :=& \mathbb{B}^{\rm MABK}_{n-1} \bigl(
                     A_1 \to A_3 ,A_2 \to A_1 ,...,B_1 \to B_3 ,B_2 \to B_1\bigr) \\
                  =&\dfrac{1}{2} \Bigl[\mathbb{B}_{n-2}^{(++)'} \bigl(C_3 +C_1\bigr)
                     +\mathbb{B}_{n-2}^{(+-)'} \bigl(C_3 -C_1\bigr)\Bigr],
               \end{align*}
               and $C_3 +C_1 =2\begin{bmatrix} 1 & 0 \\ 0 & -1 \end{bmatrix}$ is ascertained for any even $n$, then recursively, 
               \begin{equation}
                  \mathbb{B}_{n-1}^{(+-)}
                  =\begin{bmatrix}
                     1 & \cdots & 0 \\
                     \vdots & \ddots & \vdots \\
                     0 & \cdots & -1  
                  \end{bmatrix}_{2^{n-1}\times 2^{n-1}},
               \end{equation}
               which implies that
               \begin{equation*}
                  \mathbb{B}^{\rm EMABK}_n =\dfrac{1}{2} \Bigl[
                     \mathbb{B}_{n-1}^{(++)} \bigl(D_1+D_2\bigr)
                     +\mathbb{B}_{n-1}^{(+-)}\bigl(D_1-D_2\bigr)\Bigr]
                  =\begin{bmatrix}
                     \cos\theta_{d1} & \cdots & -\dfrac{1}{2^{(n-2)/2}}\sin\theta_{d1} \\
                     \vdots & \ddots & \vdots \\
                     -\dfrac{1}{2^{(n-2)/2}}\sin\theta_{d1} & \cdots & \cos\theta_{d1}  
                  \end{bmatrix}_{2^n\times 2^n},
               \end{equation*}
               and the expectation value of $\mathbb{B}^{\rm EMABK}_n$ for the $n$-qubit generalized GHZ state 
               $\cos\theta\,|00\cdots 0\rangle+\sin\theta\,|11\cdots 1\rangle$, reads 
               \begin{equation}
                  \exv{\mathbb{B}^{\rm EMABK}_n}=\cos\theta_{d1} -\dfrac{\sin(2\theta)}{2^{(n-2)/2}}\sin\theta_{d1},
               \end{equation} 
               $n>2$, and $n$ is even.

               In summary, $\forall$ even $n$, the preceding measurement settings 
               \eqref{eq:MaTheta} reduce $\exv{\mathbb{B}^{\rm EMABK}_n}$ to
               \begin{equation}
                  \begin{split}
                     \exv{\mathbb{B}^{\rm EMABK}_n} =& \cos\theta_{d1} 
                        -\dfrac{\sin(2\,\theta)}{2^{(n-2)/2}} \sin\theta_{d1}\\
                     =& \sqrt{1+\dfrac{\sin^2(2\theta)}{2^{n-2}}} \cos\bigl(
                        \theta_{d1}+\xi\bigr),
                  \end{split}
               \end{equation}
               where $\tan\xi:=\sin(2\theta)/\bigl[2^{(n-2)/2}\bigr]$. 
               $\exv{\mathbb{B}^{\rm EMABK}_n}$ is greater than 1 $\forall\,\theta\in(0,\pi/2)$, after designating $\theta_{d1}$ as some typical values, e.g. $-\xi$.
            \end{proof}
      \subsection{Proof for Odd $n$}
         \subsubsection{Lemma 1}
            In the case of odd $n$, we consider the constraints in 
            \eqref{eq:AlThetPiO2}, then $\exv{\mathbb{B}^{\rm EMABK}_n}\le 2^{(n-1)/2}$. Once 
            \begin{equation}\label{eq:PosPhiOddN}
               \begin{split}
                  & \varphi_{a1} =\varphi_{a4}=0,\ \varphi_{a2} =\varphi_{a3}
                     =\dfrac{\pi}{2}, \\
                  & \varphi_{b1}=\varphi_{b4}=\varphi_{c1}=\varphi_{c4} =\cdots
                     =\varphi_{d1} =\varphi_{d4} =\varphi_{e1} =-\dfrac{\pi}{4},\\
                  & \varphi_{b2} =\varphi_{b3} =\varphi_{c2} =\varphi_{c3} =\cdots
                     =\varphi_{d2} =\varphi_{d3} =\varphi_{e2} =\dfrac{\pi}{4},
               \end{split}
            \end{equation}
            the maximal violations are attained.
            \begin{proof}
               In comparision to the transformation rule of $\mathbb{B}^{(+-)}$ between odd and even $n$, namely
               \begin{equation*}
                  \mathbb{B}_{n-1}^{(+-)} =\begin{cases}
                     & \mathbb{B}_{n-1}^{(++)} (A_1 \to A_3 ,A_2 \to A_1,
                           B_1 \to B_3 ,B_2 \to B_1 ,...,C_1 \to C_3 ,C_2 \to C_1),\ 
                        \text{even}\ n,\\
                     & \mathbb{B}_{n-1}^{(++)} (A_1 \to A_3 ,A_2 \to A_4,
                              B_1 \to B_3 ,B_2 \to B_4 ,...,D_1 \to D_3 ,D_2 \to D_4),\ 
                           \text{odd}\ n.
                  \end{cases}
               \end{equation*}
               We conclude the permutation process for even $n$ can be recovered as long as $A_4 =A_1$, $B_4 =B_1$,...,$C_4 =C_1$ under the circumstance of odd $n$. Thus set $\varphi_{a4} =\varphi_{a1} =0$, and 
               $B_4 =B_1 =\cdots=C_4 =C_1 =-\pi/4$, then the matrix form of 
               $\bigl(\mathbb{B}^{\rm MABK}_{n-1}\bigr)'$ and $\mathbb{B}_{n-1}^{(+-)}$ for odd $n$ are same under the constraint \eqref{eq:PosPhiOddN}. 
               Therefore, $\exv{\mathbb{B}^{\rm EMABK}_n} =\exv{\mathbb{B}^{\rm MABK}_n}$ for the $n$-qubit GHZ state 
               $\bigl(|00\cdots 0\rangle+|11\cdots 1\rangle\bigr)/\sqrt{2}$. This ends the proof of \LemRef{lem:1} for odd $n$.
            \end{proof}
         \subsubsection{Lemma 2}
               As for \LemRef{lem:2}, we observe that once the following condition \eqref{eq:PhiThetOddN} holds, $\exv{\mathbb{B}^{\rm EMABK}_n}>1$ is satisfied in the whole entangled region $\theta\in(0,\pi/2)$. 
               \begin{equation}\label{eq:PhiThetOddN}
                  \begin{split}
                     & \theta_{a1} =\theta_{a2} =\theta_{b1} =\theta_{b2} =\cdots
                           =\theta_{d1} =\theta_{d2} =0;\,
                        \theta_{a3} =\theta_{a4} =\theta_{b3} =\theta_{b4} =\cdots
                           =\theta_{d3} =\theta_{d4}=\dfrac{\pi}{2},\\
                     & \varphi_{a1} =\varphi_{a2} =\varphi_{a3} =\varphi_{b1} 
                        =\varphi_{b2} =\varphi_{b3} =\cdots
                        =\varphi_{d1} =\varphi_{d2} =\varphi_{d3} =0;\,
                        \varphi_{a4} =\varphi_{b4} =\cdots=\varphi_{d4} =\dfrac{\pi}{2},
                        \\
                     & \theta_{e2}=-\theta_{e1},\ \varphi_{e2} =\varphi_{e1}
                        =\Bigl[(-1)^{(n-1)/2}\Bigr]\dfrac{\pi}{4},
                  \end{split}
               \end{equation}
               namely
               \begin{equation}\label{eq:MatOdd}
                  \begin{split}
                     & A_1 =A_2 =B_1 =B_2 =\cdots=D_1 =D_2 =\begin{bmatrix}
                        1 & 0 \\ 0 & -1 \end{bmatrix},\ 
                     A_3 =B_3 =\cdots=D_3 =\begin{bmatrix}
                        0 & 1 \\ 1 & 0 \end{bmatrix}, \\
                     & A_4 =B_4 =\cdots=D_4 =\begin{bmatrix}
                        0 & -\rm i \\ \rm i & 0 \end{bmatrix}, \\
                     & E_1 =\begin{bmatrix} \cos\theta_{e1} 
                        & \dfrac{1}{\sqrt{2}} \sin\theta_{e1} \Bigl[1-{\rm i}\,(-1)^{
                           (n-1)/2}\Bigr] \\
                        \dfrac{1}{\sqrt{2}} \sin\theta_{e1} \Bigl[1+{\rm i}\,(-1)^{
                           (n-1)/2}\Bigr] 
                        & -\cos\theta_{e1} \end{bmatrix}, \\
                     & E_2 =-\begin{bmatrix} -\cos\theta_{e1} 
                        & \dfrac{1}{\sqrt{2}} \sin\theta_{e1} \Bigl\{
                              1-\bigl[(-1)^{{(n-1)/2}}\bigr]\rm i\Bigr\} \\
                           \dfrac{1}{\sqrt{2}} \sin\theta_{e1} \Bigl\{
                              1+\bigl[(-1)^{{(n-1)/2}}\bigr]\rm i\Bigr\} 
                        & \cos\theta_{e1} \end{bmatrix}.
                  \end{split}
               \end{equation}
               \begin{proof}
                  If $n=3$,  
                  \begin{equation*}
                     \begin{split}
                        &A_1=A_2=B_1=B_2=\begin{bmatrix}1 & 0 \\ 0 & -1\end{bmatrix},\ 
                        A_3=B_3=\begin{bmatrix}0 & 1 \\ 1 & 0\end{bmatrix},\ 
                        A_4=B_4=\begin{bmatrix}0 & -\rm i \\ \rm i & 0\end{bmatrix},\\
                        &C_1=\begin{bmatrix}\cos\theta_{c1} & \dfrac{\sin\theta_{c1}(1+{\rm i})}{\sqrt{2}} \\
                           \dfrac{\sin\theta_{c1}(1-{\rm i})}{\sqrt{2}} & -\cos\theta_{c1}\end{bmatrix},\
                        C_2=-\begin{bmatrix}-\cos\theta_{c1} & \dfrac{\sin\theta_{c1}(1+{\rm i})}{\sqrt{2}} \\
                           \dfrac{\sin\theta_{c1}(1-{\rm i})}{\sqrt{2}} & \cos\theta_{c1}\end{bmatrix},
                     \end{split}
                  \end{equation*}
                  then
                  \begin{equation*}
                     C_1+C_2=2\begin{bmatrix}\cos\theta_{c1} & 0 \\ 0 & -\cos\theta_{c1}\end{bmatrix}, \
                     C_1-C_2=\sqrt{2}\begin{bmatrix}0 & \sin\theta_{c1}(1+{\rm i})\\ \sin\theta_{c1}(1-{\rm i}) & 0\end{bmatrix},
                  \end{equation*}
                  and
                  \begin{equation}
                     \mathbb{B}_2^{(++)} ={\rm CHSH}:=\dfrac{1}{2} \Bigl[A_1(B_1 +B_2)
                        +A_2(B_1-B_2)\Bigr]={\rm Diag}(1,-1,-1,1),
                  \end{equation}
                  similarly,
                  \begin{equation}
                     \mathbb{B}_2^{(+-)} :=\dfrac{1}{2} \Bigl[A_3(B_3 +B_4)
                        +A_4(B_3-B_4)\Bigr]={\rm Adiag}(1-\rm i,0,0,1+{\rm i}),
                  \end{equation}
                  which signifies that
                  \begin{equation}
                     \begin{split}
                        \mathbb{B}_3 &:= \dfrac{1}{2} \Bigl[
                           \mathbb{B}_2^{(++)} (C_1 +C_2) 
                           +\mathbb{B}_2^{(+-)} (C_1 -C_2)\Bigr] \\
                        &= {\rm Diag}(1,-1,-1,1)\otimes{\rm Diag}(\cos\theta_{c1},
                              -\cos\theta_{c1})
                           +\dfrac{\sin\theta_{c1}}{\sqrt{2}} {\rm Adiag}(
                              1-\rm i,0,0,1+{\rm i})\otimes{\rm Adiag}(
                                 1+\rm i,0,0,1-{\rm i}) \\
                        &= \cos\theta_{c1}\,{\rm Diag}(1,...,-1)
                           +\sqrt{2}\,\sin\theta_{c1}\,{\rm Adiag}(1,...,1) \\
                        &=\begin{bmatrix}
                           \cos\theta_{c1} & \cdots & \sqrt{2}\,\sin\theta_{c1}\\
                           \vdots & \ddots & \vdots \\
                           \sqrt{2}\,\sin\theta_{c1} & \cdots & -\cos\theta_{c1}  
                        \end{bmatrix}_{8\times 8}.
                     \end{split}
                  \end{equation}
                  Further, the mean value of $\mathbb{B}_3$ for the 3-qubit generalized GHZ state 
                  $\cos\theta\,|000\rangle+\sin\theta\,|111\rangle$ reads 
                  \begin{equation}  
                     \exv{\mathbb{B}_3}=\cos(2\theta)\cos\theta_{c1}
                        +\sqrt{2}\,\sin(2\theta)\sin\theta_{c1}.         
                  \end{equation}
                  
                  Once $n=5$, select the following measurement
                  \begin{equation*}
                     \begin{split}
                        & A_1 =A_2 =B_1 =B_2 =C_1 =C_2 =D_1 =D_2 =\begin{bmatrix}
                           1 & 0 \\ 0 & -1\end{bmatrix}, \\
                        & A_3 =B_3 =C_3 =D_3 =\begin{bmatrix}
                              0 & 1 \\ 1 & 0\end{bmatrix},\ 
                           A_4 =B_4 =C_3 =D_3 =\begin{bmatrix}
                              0 & -\rm i \\ \rm i & 0 \end{bmatrix}, \\
                        & E_1 =\begin{bmatrix} \cos\theta_{e1} 
                           & \dfrac{1}{\sqrt{2}} \sin\theta_{e1} (1-{\rm i}) \\
                           \dfrac{1}{\sqrt{2}} \sin\theta_{e1} (1+{\rm i})
                           & -\cos\theta_{e1} \end{bmatrix},\
                        E_2 =-\begin{bmatrix} -\cos\theta_{e1} 
                           & \dfrac{1}{\sqrt{2}} \sin\theta_{e1} (1-{\rm i}) \\
                           \dfrac{1}{\sqrt{2}} \sin\theta_{e1} (1+{\rm i}) 
                           & \cos\theta_{e1} \end{bmatrix},
                     \end{split}
                  \end{equation*}
                  then we have
                  \begin{equation}  
                     \exv{\mathbb{B}_5} =\cos(2\theta)\cos\theta_{e1} 
                        -2\sqrt{2}\,\sin(2\theta)\sin\theta_{e1}.       
                  \end{equation}
                  
                  For arbitrary odd $n$ ($n>5$),  via iteration,
                  \begin{align*}
                     & \mathbb{B}_2^{(++)} =\dfrac{1}{2} \Bigl[A_1 (B_1 +B_2)
                        +A_2 (B_1 -B_2)\Bigr]=A_1 B_1
                     =\begin{bmatrix} 1 & 0 \\ 0 & -1 \end{bmatrix}\otimes
                        \begin{bmatrix} 1 & 0 \\ 0 & -1 \end{bmatrix}
                     =\begin{bmatrix}
                           1 & \cdots & 0 \\
                           \vdots & \ddots & \vdots \\
                           0 & \cdots & 1  
                        \end{bmatrix}_{4\times 4},\\
                     & \mathbb{B}_4^{(++)} =\dfrac{1}{2} \Bigl[\mathbb{B}_3^{(++)} (
                           D_1 +D_2)+\mathbb{B}_3^{(+-)} (D_1 -D_2)\Bigr]
                        =\mathbb{B}_3^{(++)} D_1 ={\rm Diag}(1,...,-1)_{8\times 8}
                           \otimes{\rm Diag}(1,-1)
                        ={\rm Diag}(1,...,1)_{16\times 16},
                  \end{align*}
                  which means that
                  \begin{equation}
                     \mathbb{B}_{n-1}^{(++)} =\mathbb{B}_{n-1}^{\rm{MABK}} 
                     ={\rm Diag}(1,...,1)_{2^{n-1} \times 2^{n-1}},
                  \end{equation}
                  for odd $n$ under the constraints of \eqref{eq:MatOdd}.
                  
                  Likewise, 
                  \begin{align*}
                     \mathbb{B}_2^{(+-)} &=\dfrac{1}{2}\Bigl[A_3(B_3+B_4)+A_4(B_3-B_4)\Bigr]={\rm Adiag}(1-\rm i,0,0,1+{\rm i})\\
                     \mathbb{B}_4^{(+-)} &= \mathbb{B}_4^{(++)} (A_1 \to A_3,
                        A_2 \to A_4 ,...,D_1 \to D_3 ,D_2 \to D_4) \\
                     &= 2\,{\rm Adiag}(-1-\rm i,...,-1+{\rm i})_{16\times 16},\\
                     ......                        
                  \end{align*}
                  which indicates that
                  \begin{equation}
                     \mathbb{B}_{n-1}^{(+-)}=2^{(n-3)/2} \times
                     \begin{cases}
                        &{\rm Adiag}(1-\rm i,...,1+{\rm i})_{2^{n-1}\times2^{n-1}},\ \left(\dfrac{n-1}{2}\right)\,\text{is odd},\\
                        &{\rm Adiag}(-1-\rm i,...,-1+{\rm i})_{2^{n-1}\times2^{n-1}},\ \left(\dfrac{n-1}{2}\right)\,\text{is even}.
                     \end{cases}
                  \end{equation}
                  After that, we obtain
                  \begin{align*}
                     \mathbb{B}^{\rm EMABK}_n&=\dfrac{1}{2}\Bigl[
                        \mathbb{B}_{n-1}^{(++)}\bigl(E_1+E_2\bigr)
                        +\mathbb{B}_{n-1}^{(+-)}\bigl(E_1-E_2\bigr)\Bigr]\\
                     &=\begin{cases}
                        & \begin{bmatrix}
                           \cos\theta_{e1} & \cdots & 2^{(n-2)/2}\sin\theta_{e1} \\
                           \vdots & \ddots & \vdots \\
                           2^{(n-2)/2}\sin\theta_{e1} & \cdots & -\cos\theta_{e1}  
                        \end{bmatrix}_{2^n\times 2^n},\ \left(\dfrac{n-1}{2}\right)\,\text{is odd}, \\
                        & \\
                        & \begin{bmatrix}
                           \cos\theta_{e1} & \cdots & -2^{(n-2)/2}\sin\theta_{e1} \\
                           \vdots & \ddots & \vdots \\
                           -2^{(n-2)/2}\sin\theta_{e1} & \cdots & -\cos\theta_{e1}  
                        \end{bmatrix}_{2^n\times 2^n},\ \left(\dfrac{n-1}{2}\right)\,\text{is even},
                     \end{cases}
                  \end{align*}
                  and the expectation value of $\mathbb{B}^{\rm EMABK}_n$ for the $n$-qubit generalized GHZ state 
                  $\cos\theta|00\cdots 0\rangle+\sin\theta|11\cdots 1\rangle$, reads 
                  \begin{equation}
                     \exv{\mathbb{B}^{\rm EMABK}_n}=\cos(2\theta)\cos\theta_{e1}
                        +\bigl[(-1)^{{(n+1)/2}}\bigr]2^{(n-2)/2}\sin(2\theta)\sin\theta_{e1},
                  \end{equation} 
                  $n>2$, and $n$ is odd.

                  In conclusion, $\forall$ odd $n$, $n\geq 3$, under the condition of \eqref{eq:MatOdd}, the expectation value of $\mathbb{B}^{\rm EMABK}_n$ reads
                  \begin{equation}\label{eq:ExvOddIn}
                     \begin{split}
                        \exv{\mathbb{B}^{\rm EMABK}_n}=&\cos(2\theta)\cos\theta_{e1}
                           +\bigl[(-1)^{{(n+1)/2}}\bigr]2^{(n-2)/2}\sin(2\theta)\sin\theta_{e1}\\
                        =&\sqrt{\cos^2(2\theta)+2^{(n-2)}{\sin^2}(2\theta)} \cos\Bigl[
                           \theta_{e1}+(-1)^{(n-1)/2} \zeta\Bigr],
                     \end{split}
                  \end{equation}
                  where $\tan\zeta:=\bigl[(-1)^{{(n+1)/2}}\bigr]\tan(2\theta)$. Obviously, 
                  $\forall\,\theta\in(0,\pi/2),\ \exv{\mathbb{B}^{\rm EMABK}_n}>1$ for some typical valued $\theta_{e1}$, such as 
                  $\bigl[(-1)^{{(n+1)/2}}\bigr]\zeta$.
               \end{proof}
\section{\'Sliwa's 46 Inequalities and Their Tight Generalizations}
   \subsection{Decomposition of \'Sliwa's 46 Inequalities}
      In the following table, the 46 \'Sliwa's inequalities \cite{2003Sliwa} are rewritten via the iterative formula  
      \begin{align}
         \mathbb{B}_3 =\dfrac{1}{2} \Bigl[\mathbb{B}_2^{(++)} (C_1 +C_2) 
            +\mathbb{B}_2^{(+-)} (1-C_2) +\mathbb{B}_2^{(-+)} (1-C_1)\Bigl]\le 1,
      \end{align}
      under the condition
      \begin{equation}
         \mathbb{B}_2^{(--)} =\mathbb{B}_2^{(+-)} +\mathbb{B}_2^{(-+)} 
            -\mathbb{B}_2^{(++)}.
      \end{equation}
\begin{longtable}[htbp]{|c|*3{p{5.2cm}|}}
         \caption{46 Normalized \'Sliwa's inequalities. $\text{\'Sliwa}_l$ indicates 
            the $l$-th \'Sliwa's inequality.} \\
         \hline
         Number & \multicolumn{3}{|c|}{Bell Inequalities} \\
         \hline\hline
         $\text{\'Sliwa}_1$ &  \multicolumn{3}{|c|}{$A_1 +B_1 +C_1 -A_1 C_1 
            -B_1 C_1 -A_1 B_1 +A_1B_1C_1\le 1$} \\
         \hline
         Split Form & $\mathbb{B}_2^{(++)}$ & $\mathbb{B}_2^{(+-)}$ 
            & $\mathbb{B}_2^{(-+)}$ \\
         \hline
         & 1 & 1 & $-1+2\,A_1 +2\,B_1 -2\,A_1 B_1$  \\ 
         \hline\hline
         $\text{\'Sliwa}_2$ & \multicolumn{3}{|c|}{$\dfrac{1}{2} (A_1 B_1 C_1 +A_2 B_2 C_1
            +A_2 B_1 C_2 -A_1 B_2 C_2)\le 1$} \\
         \hline
         Split Form & $\mathbb{B}_2^{(++)}$ & $\mathbb{B}_2^{(+-)}$ 
            & $\mathbb{B}_2^{(-+)}$ \\
         \hline
         & $\dfrac{1}{2} (A_1 B_1 +A_2 B_1 -A_1 B_2 +A_2 B_2)$  
            & $\dfrac{1}{2} (A_1 B_1 -A_2 B_1 +A_1 B_2 +A_2 B_2)$  
            & $-\mathbb{B}_2^{(+-)}$  \\
\hline
      \end{longtable}
   \subsection{A Family of $(4,2,2)$ Inequalities}
      Next we will start from \'Sliwa's 46 tight $(3,2,2)$ inequalities 
      \cite{2003Sliwa}, adopting the equivalent transformations and iterative formula 
      \begin{align}
         \mathbb{B}_4 =\dfrac{1}{2} \Bigl[\mathbb{B}_3^{(++)} (D_1 +D_2) 
            +\mathbb{B}_3^{(+-)} (1-D_2) +\mathbb{B}_3^{(-+)} (1-D_1)\Bigl]\le 1,
      \end{align}
      under the condition
      \begin{equation}
         \mathbb{B}_3^{(--)} =\mathbb{B}_3^{(+-)} +\mathbb{B}_3^{(-+)} 
            -\mathbb{B}_3^{(++)},
      \end{equation}
      to establish tight $(4,2,2)$ Bell inequalities. For simplicity, all 
      $\mathbb{B}_3^{(++)}$'s are set to associated \'Sliwa's inequalities after normalization, which are positioned in the first row of every table below. Note that all newly generated inequalities are normalized and $\mathcal{Q}$ refers to the numerical quantum upper bound of corresponding inequality. Whereas, the new 
      $(4,2,2)$ inequalities are too cumbersome to be listed completely after 
      $\text{\'Sliwa}_5$, so we present two of them merely and list them entirely in \cite{NewTightBI}, which is open source.
\begin{longtable}[htbp]{|c|*2{p{3cm}|}}            
         \caption{The (4,2,2) tight inequality generated from $\text{\'Sliwa}_1$.}
            \label{tab:Tigh4PSliwa1} \\
         \hline
         $\text{\'Sliwa}_1$ & \multicolumn{1}{|c|}{$A_1+B_1-A_1 B_1+C_1-A_1 C_1-B_1 C_1
               +A_1 B_1 C_1\le 1$} 
            & \multicolumn{1}{|c|}{$\mathcal{Q}=1$} \\
         \hline\hline            
         Number & \multicolumn{1}{|c|}{New Tight Inequality} & \multicolumn{1}{|c|}{
               Remarks} \\
         \hline
         1 & \multicolumn{1}{|p{9cm}|}{$\dfrac{1}{2}\bigl(A_1 B_1 C_1 D_1
               +A_1 B_1 C_2 D_1 +A_1 B_1 C_1 D_2 -A_1 B_1 C_2 D_2-2A_1 B_1-A_1 C_1 D_1
               -A_1 C_2 D_1-A_1 C_1 D_2+A_1 C_2 D_2+2A_1-B_1 C_1 D_1-B_1 C_2 D_1
               -B_1 C_1 D_2+B_1 C_2 D_2+2 B_1+C_1 D_1+C_2 D_1+C_1 D_2-C_2 D_2\bigr)
               \le 1$} 
            & \multicolumn{1}{|c|}{$\mathcal{Q}=4\sqrt{2}-3$} \\
         \hline
         $\mathbb{B}_3^{(+-)}$ & \multicolumn{1}{|c|}{$A_1+B_1-A_1 B_1+C_2-A_1 C_2
               -B_1 C_2+A_1 B_1 C_2\le 1$} 
            & \multicolumn{1}{|c|}{$\mathbb{B}_3^{(++)}\overset{C_{1\leftrightarrow 2}}
               {\longleftrightarrow}\mathbb{B}_3^{(+-)}$} \\
         \hline
         $\mathbb{B}_3^{(-+)}$ & \multicolumn{1}{|c|}{$A_1+B_1-A_1 B_1-C_2+A_1 C_2
               +B_1 C_2-A_1 B_1 C_2\le 1$} 
            & \multicolumn{1}{|c|}{$\mathbb{B}_3^{(++)}\overset{C_{1\leftrightarrow 2},
               C_2\to -C_2}{\longleftrightarrow}\mathbb{B}_3^{(-+)}$} \\
         \hline
         $\mathbb{B}_3^{(--)}$ & \multicolumn{1}{|c|}{$A_1+B_1-A_1 B_1-C_1+A_1 C_1
            +B_1 C_1 -A_1 B_1 C_1\le 1$} & \multicolumn{1}{|c|}{$\mathbb{B}_3^{(++)} \overset{C_1 \to-C_1}{\longleftrightarrow}\mathbb{B}_3^{(--)}$} \\        
         \hline         
      \end{longtable}
      \begin{longtable}[htbp]{|c|*2{p{3cm}|}}            
         \caption{The (4,2,2) tight inequalities generated from $\text{\'Sliwa}_2$.}\label{tab:Tigh4PSliwa2} \\
         \hline
         $\text{\'Sliwa}_2$ & \multicolumn{1}{|c|}{$\dfrac{1}{2}\bigl(A_1 B_1 C_1
               +A_2 B_2 C_1+A_2 B_1 C_2-A_1 B_2 C_2\bigr)\le 1$} 
            & \multicolumn{1}{|c|}{$\mathcal{Q}=2$} \\
         \hline\hline          
         Number & \multicolumn{1}{|c|}{New Tight Inequalities} & \multicolumn{1}{|c|}{
            Remarks} \\
         \hline
         1 & \multicolumn{1}{|c|}{$\mathbb{B}_3^{(++)}\overset{A_1\to-A_1}{
               \longleftrightarrow}\mathbb{B}_3^{(+-)},\ 
            \mathbb{B}_3^{(++)}\overset{A_2\to-A_2}{
               \longleftrightarrow}\mathbb{B}_3^{(-+)},\ 
            \mathbb{B}_3^{(++)}\overset{A\leftrightarrow C,B_1\to-B_1}{
               \longleftrightarrow}\mathbb{B}_3^{(--)}$} 
            & \multicolumn{1}{|c|}{$\mathbb{B}_3^{(-+)}=-\mathbb{B}_3^{(+-)}$,\ 
               $\mathcal{Q}=2$} \\       
         \hline
         2 & \multicolumn{1}{|c|}{$\mathbb{B}_3^{(++)}\overset{C_{1\leftrightarrow 2}}{\longleftrightarrow}\mathbb{B}_3^{(+-)},\ 
            \mathbb{B}_3^{(++)}\overset{A_{1\leftrightarrow 2},B_1\to-B_1}{\longleftrightarrow}\mathbb{B}_3^{(-+)},\ 
               \mathbb{B}_3^{(++)}\overset{A\leftrightarrow C,B_1\to-B_1}{\longleftrightarrow}\mathbb{B}_3^{(--)}$} 
            & \multicolumn{1}{|c|}{$\mathbb{B}_3^{(-+)}=-\mathbb{B}_3^{(+-)}$,\ 
               $\mathcal{Q}=2$} \\
\hline                     
      \end{longtable}        
      \begin{longtable}[htbp]{|c|*2{p{2cm}|}}            
         \caption{The (4,2,2) tight inequalities generated from $\text{\'Sliwa}_3$.}
            \label{tab:Tigh4PSliwa3} \\
         \hline
         $\text{\'Sliwa}_3$ & \multicolumn{1}{|c|}{$\dfrac{1}{2}\bigl(A_1 B_1 C_1
               +A_2 B_1 C_1 +A_1 B_2 C_2-A_2 B_2 C_2\bigr)\le 1$}
            & \multicolumn{1}{|c|}{$\mathcal{Q}=\sqrt{2}$} \\
         \hline\hline          
         Number & \multicolumn{1}{|c|}{New Tight Inequalities} 
            & \multicolumn{1}{|c|}{Remarks} \\
         \hline
         1 & \multicolumn{1}{|c|}{$\mathbb{B}_3^{(++)}\overset{C_{1\leftrightarrow 2}}{
               \longleftrightarrow}\mathbb{B}_3^{(+-)},\ 
            \mathbb{B}_3^{(++)}\overset{A_1\to-A_1,B_{1\leftrightarrow 2}}{
               \longleftrightarrow}\mathbb{B}_3^{(-+)},\
            \mathbb{B}_3^{(++)}\overset{B_1\to-B_1,C_2\to-C_2}
               {\longleftrightarrow}\mathbb{B}_3^{(--)}$} 
            & \multicolumn{1}{|c|}{$\mathbb{B}_3^{(-+)}=-\mathbb{B}_3^{(+-)}$,\ 
               $\mathcal{Q}=\sqrt{2}$} \\        
         \hline
         2 & \multicolumn{1}{|c|}{$\mathbb{B}_3^{(++)}\overset{A\leftrightarrow B}{
               \longleftrightarrow}\mathbb{B}_3^{(+-)},\ 
            \mathbb{B}_3^{(++)}\overset{A_1\to-A_1,B_{1\leftrightarrow 2}}{
               \longleftrightarrow}\mathbb{B}_3^{(-+)},\ 
            \mathbb{B}_3^{(++)}\overset{A_1\to-A_1,B_{1\leftrightarrow 2},
               A\leftrightarrow B}{\longleftrightarrow}
            \mathbb{B}_3^{(--)}$} & \multicolumn{1}{|c|}{$\mathcal{Q}=\sqrt{2}$} \\
\hline                                       
      \end{longtable}
      \begin{longtable}[htbp]{|c|*2{p{2cm}|}}            
         \caption{The (4,2,2) tight inequalities generated from $\text{\'Sliwa}_4$.} \\
         \hline
         $\text{\'Sliwa}_4$ & \multicolumn{1}{|p{11cm}|}{$\dfrac{1}{2}\bigl(2A_1
            +B_1 C_1-A_1 B_1 C_1+B_2 C_1-A_1 B_2 C_1+B_1 C_2 -A_1 B_1 C_2-B_2 C_2
            +A_1 B_2 C_2\bigr)\le 1$} & \multicolumn{1}{|c|}{$\mathcal{Q}=2\sqrt{2}-1$}
            \\
         \hline\hline          
         Number & \multicolumn{1}{|c|}{New Tight Inequalities} & \multicolumn{1}{|c|}{
            Remarks} \\
         \hline\hline
         1 & \multicolumn{1}{|c|}{$\mathbb{B}_3^{(++)}\overset{B_1\to-B_1}
               {\longleftrightarrow}\mathbb{B}_3^{(+-)},\ 
            \mathbb{B}_3^{(++)}\overset{C_{1\leftrightarrow 2}}{\longleftrightarrow}
               \mathbb{B}_3,\ 
            \mathbb{B}_3^{(++)}\overset{B_1\to-B_1,C_{1\leftrightarrow 2}}
               {\longleftrightarrow}\mathbb{B}_3^{(--)}$} 
            & \multicolumn{1}{|c|}{$\mathcal{Q}=2\sqrt{2}-1$} \\                
         \hline\hline
         2 & \multicolumn{1}{|p{9cm}|}{$\dfrac{1}{2}\bigl(-A_ 1 B_1 C_ 1 D_ 1
               +A_ 1 B_ 2 C_ 2 D_ 1-A_ 1 B_ 2 C_1 D_ 2 -A_ 1 B_ 1 C_ 2 D_ 2+2 A_ 1
               +B_ 1 C_ 1 D_ 1-B_ 2 C_ 2 D_ 1+B_ 2 C_ 1 D_ 2+B_ 1 C_ 2 D_ 2\bigr)
               \le 1$} 
            & \multicolumn{1}{|c|}{$\mathcal{Q}=3$} \\
         \hline
         $\mathbb{B}_3^{(+-)}$ & \multicolumn{1}{|p{9cm}|}{$\dfrac{1}{2}\bigl(2A_1
               +B_1 C_1-A_1 B_1 C_1-B_2 C_1+A_1 B_2 C_1-B_1 C_2 +A_1 B_1 C_2-B_2 C_2
               +A_1 B_2 C_2\bigr)\le 1$}  
            & \multicolumn{1}{|c|}{$\mathbb{B}_3^{(++)}\overset{B_1\to-B_1,
               B_{1\leftrightarrow 2}}{\longleftrightarrow}\mathbb{B}_3^{(+-)}$} \\
\hline   
      \end{longtable}
      \begin{longtable}[htbp]{|c|*2{p{2.5cm}|}}            
         \caption{The (4,2,2) tight inequalities generated from $\text{\'Sliwa}_5$.}\\
         \hline
         $\text{\'Sliwa}_5$ & \multicolumn{1}{|p{9cm}|}{$\dfrac{1}{3}\bigl(A_1+B_1
               +A_2 B_1+A_1 B_2-A_2 B_2+C_1+A_2 C_1 -A_1 B_1 C_1-A_2 B_1 C_1+B_2 C_1
               -A_1 B_2 C_1+A_1 C_2-A_2 C_2+B_1 C_2-A_1 B_1 C_2-B_2 C_2+A_2 B_2 C_2
               \bigr)\le 1$}
            & \multicolumn{1}{|c|}{$\mathcal{Q}=\dfrac{8\sqrt{5}-13}{3}$} \\
         \hline\hline          
         Number & \multicolumn{1}{|c|}{New Tight Inequalities} & \multicolumn{1}{|c|}{
            Remarks} \\
         \hline\hline
         1 & \multicolumn{1}{|p{9cm}|}{$\dfrac{1}{6}\bigl(-2A_ 1 B_ 1 C_ 1 D_ 1
               -A_ 1 B_ 1 C_2 D_ 1+A_2 B_ 1 C_ 2 D_ 1 -2 A_ 2 B_ 1 C_ 1 D_ 2
               -A_ 1 B_ 1 C_ 2 D_ 2-A_ 2 B_ 1 C_ 2 D_ 2-A_ 1 B_ 2 C_ 1 D_ 1
               +A_ 2 B_ 2 C_ 1 D_ 1 +A_ 1 B_ 2 C_ 2 D_ 1+A_ 2 B_ 2 C_ 2 D_ 1
               -A_ 1 B_ 2 C_ 1 D_ 2-A_ 2 B_ 2 C_ 1 D_ 2-A_ 1 B_ 2 C_ 2 D_ 2 
               +A_ 2 B_ 2 C_ 2 D_ 2+A_ 1 B_ 1 D_ 1+A_ 2 B_ 1 D_ 1-A_ 1 B_ 1 D_ 2
               +A_ 2 B_ 1 D_ 2-2 A_2 B_ 2 D_ 1 +2A_ 1 B_ 2 D_ 2+A_ 1 C_ 1 D_ 1
               +A_ 2 C_ 1 D_ 1-2 A_ 2 C_2 D_ 1-A_ 1 C_ 1 D_ 2+A_ 2 C_ 1 D_ 2 
               +2 A_ 1 C_ 2 D_ 2+A_ 1 D_ 1-A_2 D_ 1+A_ 1 D_ 2+A_ 2 D_ 2+2 B_ 1 C_ 2
               +2 B_ 2 C_ 1-2 B_ 2 C_ 2+2 B_ 1+2C_1\bigr)\le 1$} 
            & \multicolumn{1}{|c|}{$\mathcal{Q}=2.41$} \\
         \hline
         $\mathbb{B}_3^{(+-)}$ & \multicolumn{1}{|p{9cm}|}{$\dfrac{1}{3}\bigl(-A_2
               +B_1+A_1 B_1-A_1 B_2-A_2 B_2+C_1+A_1 C_1 -A_1 B_1 C_1+A_2 B_1 C_1
               +B_2 C_1+A_2 B_2 C_1-A_1 C_2-A_2 C_2+B_1 C_2+A_2 B_1 C_2-B_2 C_2
               +A_1 B_2 C_2\bigr)\le 1$} 
            & \multicolumn{1}{|c|}{$\mathbb{B}_3^{(++)}\overset{A_2\to-A_2,
               B_1\to-B_1}{\longleftrightarrow}\mathbb{B}_3^{(+-)}$} \\
\hline
      \end{longtable}
      \begin{longtable}[htbp]{|c|*2{p{3cm}|}}            
         \caption{Two cases of the (4,2,2) tight inequalities generated from 
            $\text{\'Sliwa}_6$.} \\
         \hline
         $\text{\'Sliwa}_6$ & \multicolumn{1}{|p{9cm}|}{$\dfrac{1}{3}\bigl(A_1+B_1
               +A_1 B_1+C_1+A_2 C_1-A_1 B_1 C_1 -A_2 B_1 C_1+B_2 C_1-A_1 B_2 C_1
               +A_1 C_2-A_2 C_2-B_1 C_2+A_2 B_1 C_2+B_2 C_2-A_1 B_2 C_2\bigr)\le 1$}
            & \multicolumn{1}{|c|}{$\mathcal{Q}=\dfrac{4\sqrt{2}-1}{3}$} \\
         \hline\hline          
         Number & \multicolumn{1}{|c|}{New Tight Inequalities} & \multicolumn{1}{|c|}{
            Remarks} \\
         \hline
         1 & \multicolumn{1}{|c|}{$\mathbb{B}_3^{(++)}\overset{A\leftrightarrow C,
                  B_1\to-B_1,B_2\to-B_2}{\longleftrightarrow}
                     \mathbb{B}_3^{(+-)},\ 
               \mathbb{B}_3^{(++)}\overset{A\leftrightarrow C}
                  {\longleftrightarrow}\mathbb{B}_3^{(-+)},\ 
               \mathbb{B}_3^{(++)}\overset{B_1\to-B_1,
                  B_2\to-B_2}{\longleftrightarrow}\mathbb{B}_3^{(--)}$} 
            & \multicolumn{1}{|c|}{$\mathcal{Q}=\dfrac{4\sqrt{2}-1}{3}$} \\        
         \hline
         2 & \multicolumn{1}{|p{9cm}|}{$\dfrac{1}{6}\bigl(-2 A_ 1 B_ 1 C_ 1 D_ 1
               -2 A_ 1 B_ 2 C_ 1 D_ 1 -A_ 1 B_ 1 C_ 2 D_ 1-A_ 1 B_ 2 C_ 2 D_ 1
               +A_ 1 B_ 1 C_ 2 D_ 2-A_ 1 B_ 2 C_ 2 D_ 2-A_ 2 B_ 1 C_ 1 D_ 1
               -A_ 2 B_ 2 C_ 1 D_ 1+2 A_ 2 B_ 1 C_ 2 D_ 1-A_ 2 B_ 1 C_ 1 D_ 2
               +A_2 B_ 2 C_ 1 D_ 2+A_ 1 B_ 1 D_ 1 +A_ 1 B_ 2 D_ 1+A_ 1 B_ 1 D_ 2
               -A_ 1 B_2 D_ 2-A_ 2 B_ 1 D_ 1+A_ 2 B_ 2 D_ 1+A_ 2 B_ 1 D_ 2
               -A_ 2 B_ 2 D_ 2 +2A_ 1 C_ 2+2 A_ 2 C_ 1-2 A_ 2 C_ 2+2 A_ 1
               +B_ 1 C_ 1 D_ 1+B_ 2 C_ 1 D_1-B_ 1 C_ 2 D_ 1+B_ 2 C_ 2 D_ 1 -B_ 1 C_ 1 D_ 2+B_ 2 C_ 1 D_ 2-B_ 1 C_2 D_ 2+B_ 2 C_ 2 D_ 2+2 B_ 1 D_ 1+2 C_ 1\bigr)
               \le 1$} 
            & \multicolumn{1}{|c|}{$\mathcal{Q}=2.41$} \\
\hline          
      \end{longtable}
      \begin{longtable}[htbp]{|c|*2{p{2cm}|}}            
         \caption{Two cases of the (4,2,2) tight inequalities generated from 
            $\text{\'Sliwa}_7$.} \\
         \hline
         $\text{\'Sliwa}_7$ & \multicolumn{1}{|p{9cm}|}{$\dfrac{1}{4}\bigl(
               3A_1 B_1 C_1+A_2 B_1 C_1+A_1 B_2 C_1-A_2 B_2 C_1
               +A_1 B_1 C_2-A_2 B_1 C_2-A_1 B_2 C_2+A_2 B_2 C_2\bigr)\le 1$}
            & \multicolumn{1}{|c|}{$\mathcal{Q}=\dfrac{5}{3}$} \\
         \hline\hline          
         Number & \multicolumn{1}{|c|}{New Tight Inequalities} & \multicolumn{1}{|c|}{
            Remarks} \\
         \hline
         1 & \multicolumn{1}{|p{9cm}|}{$\mathbb{B}_3^{(++)}\overset{
                  A_{1\leftrightarrow 2},B_1\to-B_1,C_1 \to-C_1}
                     {\longleftrightarrow}\mathbb{B}_3^{(+-)}$,\ 
               $\mathbb{B}_3^{(++)}\overset{B_{1\leftrightarrow 2},
                  C_{1\leftrightarrow 2},B_1\to-B_1,
                     C_2\to-C_2}{\longleftrightarrow}\mathbb{B}_3^{(-+)}$,\ 
               $\mathbb{B}_3^{(++)}\overset{A_2\to-A_2,
                  B_{1\leftrightarrow 2},C_{1\leftrightarrow 2},A_1\to-A_1,
                     A_{1\leftrightarrow 2}}{\longleftrightarrow}\mathbb{B}_3^{(--)}$} 
            & \multicolumn{1}{|c|}{$\mathcal{Q}=\dfrac{5}{3}$} \\        
         \hline
         2 & \multicolumn{1}{|p{9cm}|}{$\dfrac{1}{4}\bigl(A_ 1 B_ 1 C_ 1 D_ 1
               -A_ 2 B_ 2 C_ 1 D_ 1+A_ 1 B_ 1 C_ 1 D_ 2 +A_ 2 B_ 1 C_ 1 D_ 2
               +A_ 1 B_2 C_ 1 D_ 2+A_ 2 B_ 2 C_ 1 D_ 2-A_ 1 B_ 1 C_ 2 D_ 1
               +A_ 2 B_ 2 C_ 2 D_ 1 +A_ 1 B_ 1 C_ 2 D_ 2-A_ 2 B_ 1 C_ 2 D_ 2
               -A_ 1 B_ 2 C_ 2 D_ 2+A_ 2 B_ 2 C_ 2 D_ 2+A_ 1 B_ 1 C_ 1-A_ 2 B_ 2 C_ 1
               +A_ 1 B_ 1 C_ 2-A_ 2 B_ 2 C_ 2\bigr)\le 1$} 
            & \multicolumn{1}{|c|}{$\mathcal{Q}=2$} \\
\hline           
      \end{longtable}
      \begin{longtable}[htbp]{|c|*2{p{2cm}|}}            
         \caption{Two case of (4,2,2) tight inequalities generated from 
            $\text{\'Sliwa}_8$.} \\
         \hline
         $\text{\'Sliwa}_8$ & \multicolumn{1}{|p{9cm}|}{$\dfrac{1}{4}\bigl(A_1 B_1
               +A_2 B_1+A_1 B_2+A_2 B_2+2A_1 B_1 C_1 -2A_2 B_2 C_1+A_1 B_1 C_2
               -A_2 B_1 C_2-A_1 B_2 C_2+A_2 B_2 C_2\bigr)\le 1$}
            & \multicolumn{1}{|c|}{$\mathcal{Q}=\dfrac{5}{3}$} \\
         \hline    
         Number & \multicolumn{1}{|c|}{New Tight Inequalities} & \multicolumn{1}{|c|}{
            Remarks} \\
         \hline
         1 & \multicolumn{1}{|p{9cm}|}{$\mathbb{B}_3^{(++)}\overset{
                  B_{1\leftrightarrow 2},C_{1\leftrightarrow 2}}{\longleftrightarrow}
                     \mathbb{B}_3^{(+-)},\ 
               \mathbb{B}_3^{(++)}\overset{B\leftrightarrow C,A_{1\leftrightarrow 2},
                  B_{1\leftrightarrow 2},C_1 \to-C_1}{\longleftrightarrow}
                     \mathbb{B}_3^{(-+)}$,\ 
               $\mathbb{B}_3^{(++)}\overset{B\leftrightarrow C,B_2\to-B_2,
                  C_1 \to-C_1}{\longleftrightarrow}\mathbb{B}_3^{(--)}$} 
            & \multicolumn{1}{|c|}{$\mathcal{Q}=1.67$} \\
         \hline                    
         3 & \multicolumn{1}{|p{9cm}|}{$\dfrac{1}{8}\bigl(3 A_ 1 B_ 1 C_ 1 D_ 1
               -A_ 2 B_ 1 C_ 1 D_ 1+A_ 1 B_ 1 C_ 2 D_ 1 -3 A_ 2 B_ 1 C_ 2 D_ 1
               +A_ 1 B_ 1 C_ 1 D_ 2+A_ 2 B_ 1 C_ 1 D_ 2+A_ 1 B_ 1 C_ 2 D_ 2
               +A_ 2 B_ 1 C_ 2 D_ 2 -A_ 1 B_ 2 C_ 1 D_ 1-A_ 2 B_ 2 C_ 1 D_ 1
               +A_ 1 B_ 2 C_ 2 D_ 1+A_ 2 B_ 2 C_ 2 D_ 1+A_ 1 B_ 2 C_ 1 D_ 2
               -3 A_ 2 B_ 2 C_ 1 D_ 2-3 A_ 1 B_ 2 C_ 2 D_ 2+A_ 2 B_ 2 C_ 2 D_ 2
               +2 A_ 1 B_ 1+2 A_ 2 B_ 1+2 A_ 1 B_ 2 +2 A_ 2 B_ 2\bigr)\le 1$} 
            & \multicolumn{1}{|c|}{$\mathcal{Q}=2$} \\
\hline                    
      \end{longtable}
      \begin{longtable}[htbp]{|c|*2{p{2cm}|}}            
         \caption{Two cases (4,2,2) tight inequalities generated from 
            $\text{\'Sliwa}_9$.} \\
         \hline
         $\text{\'Sliwa}_9$ & \multicolumn{1}{|p{8cm}|}{$\dfrac{1}{4}\bigl(A_1 B_1
               +A_2 B_1+A_1 B_2+A_2 B_2+2A_1 B_1 C_1 -2A_1 B_2 C_1+A_1 B_1 C_2
               -A_2 B_1 C_2+A_1 B_2 C_2-A_2 B_2 C_2\bigr)\le 1$}
            & \multicolumn{1}{|c|}{$\mathcal{Q}=\sqrt{2}$} \\
         \hline         
         Number & \multicolumn{1}{|c|}{New Tight Inequalities} & \multicolumn{1}{|c|}{
            Remarks} \\
         \hline                               
         3 & \multicolumn{1}{|p{8cm}|}{$\dfrac{1}{8}\bigl(A_ 1 B_ 1 C_ 1 D_ 1
            +A_ 2 B_ 1 C_ 1 D_ 1-3 A_ 1 B_ 2 C_ 1 D_ 1+A_ 2 B_ 2 C_ 1 D_ 1
            +A_ 1 B_ 1 C_ 1 D_ 2-A_ 2 B_ 1 C_ 1 D_ 2-A_ 1 B_ 2 C_ 1 D_ 2
            +A_ 2 B_ 2 C_ 1 D_ 2+A_ 1 B_ 1 C_ 2 D_ 1-3 A_ 2 B_ 1 C_ 2 D_ 1
            +A_ 1 B_ 2 C_ 2 D_ 1+A_ 2 B_ 2 C_ 2 D_ 1-A_ 1 B_ 1 C_ 2 D_ 2
            +A_ 2 B_ 1 C_ 2 D_ 2+A_ 1 B_ 2 C_ 2 D_ 2-A_ 2 B_ 2 C_ 2 D_ 2
            +2 A_ 1 B_ 1 C_ 1-2 A_ 2 B_ 2 C_ 1+2 A_ 1 B_ 1 C_ 2-2 A_ 2 B_ 2 C_ 2
            +2 A_ 1 B_ 1 D_ 2+2 A_ 2 B_ 1 D_ 2+2 A_ 1 B_ 2 D_ 2+2 A_ 2 B_ 2 D_ 2
            \bigr)\le 1$} & \multicolumn{1}{|c|}{$\mathcal{Q}=2$} \\
         \hline
         $\mathbb{B}_3^{(+-)}$ & \multicolumn{1}{|p{8cm}|}{$\dfrac{1}{4}\bigl(
               A_ 1 B_ 1 C_ 1+A_ 2 B_ 1 C_ 1+2 A_ 1 B_ 1 C_ 2 -2 A_ 2 B_ 1 C_ 2
               -A_ 1 B_ 2 C_ 1-A_ 2 B_ 2 C_ 1-A_ 1 B_ 1-A_ 2 B_ 1-A_ 1 B_ 2-A_ 2 B_ 2
               \bigr)\le 1$} 
            & \multicolumn{1}{|p{6.7cm}|}{$\mathbb{B}_3^{(++)}\overset{
               A\leftrightarrow B,C_2\to-C_2,A_2\to-A_2,
               C_{1\leftrightarrow 2},A_{1\leftrightarrow 2},A_2\to-A_2}
                  {\longleftrightarrow}\mathbb{B}_3^{(+-)}$} \\
\hline                    
      \end{longtable}   
      \begin{longtable}[htbp]{|c|*2{p{2cm}|}}            
         \caption{Two cases of the (4,2,2) tight inequalities generated from 
            $\text{\'Sliwa}_{10}$.} \\
         \hline
         $\text{\'Sliwa}_{10}$ & \multicolumn{1}{|p{9cm}|}{$\dfrac{1}{4}\bigl(A_1 B_1
               +A_2 B_1+A_1 B_2+A_2 B_2+A_1 C_1-A_2 C_1+B_1 C_1+A_1 B_1 C_1-B_2 C_1
               -A_2 B_2 C_1+A_1 C_2-A_2 C_2-B_1 C_2+A_2 B_1 C_2+B_2 C_2-A_1 B_2 C_2
               \bigr)\le 1$}
            & \multicolumn{1}{|c|}{$\mathcal{Q}=1$} \\
         \hline         
         Number & \multicolumn{1}{|c|}{New Tight Inequalities} & \multicolumn{1}{|c|}{
            Remarks} \\
         \hline
         1 & \multicolumn{1}{|p{9cm}|}{$\mathbb{B}_3^{(++)}\overset{
               B_{1\leftrightarrow 2},C_{1\leftrightarrow 2}}{\longleftrightarrow}
                  \mathbb{B}_3^{(+-)}$,\newline
            $\mathbb{B}_3^{(++)}\overset{A\leftrightarrow B,A\leftrightarrow C,
               B_1\to-B_1,C_1 \to-C_1}{\longleftrightarrow}
                  \mathbb{B}_3^{(-+)}$,\newline
            $\mathbb{B}_3^{(++)}\overset{B\leftrightarrow C,A_1\to-A_1}
               {\longleftrightarrow}\mathbb{B}_3^{(--)}$} & \\
         \cline{1-2}                               
         2 & \multicolumn{1}{|p{9cm}|}{$\dfrac{1}{4}\bigl(A_ 1 B_ 1 C_ 1 D_ 2
               +A_ 2 B_ 1 C_ 2 D_ 2-A_ 2 B_ 2 C_ 1 D_ 2-A_ 1 B_ 2 C_ 2 D_ 2+A_ 1 B_ 1 D_ 1+A_ 2 B_ 2 D_ 1+A_ 2 B_ 1+A_ 1 B_ 2+A_ 1 C_ 1 D_ 1-A_ 2 C_ 2 D_ 1
               -A_ 2 C_ 1+A_ 1 C_ 2-B_ 1 C_ 2 D_ 1-B_ 2 C_ 1 D_ 1+B_ 1 C_ 1+B_ 2 C_ 2
               \bigr)\le 1$} 
            & \multicolumn{1}{|c|}{$\mathcal{Q}=\sqrt{2}$} \\
\hline                               
      \end{longtable}
      \begin{longtable}[htbp]{|c|*3{p{2cm}|}}            
         \caption{Two cases of the (4,2,2) tight inequalities generated from 
            $\text{\'Sliwa}_{11}$.} \\
         \hline
         $\text{\'Sliwa}_{11}$ & \multicolumn{1}{|p{9cm}|}{$\dfrac{1}{4}\bigl(2A_1 B_1
               +2A_2 B_2+A_1 B_1 C_1+A_2 B_1 C_1-A_1 B_2 C_1-A_2 B_2 C_1+A_1 B_1 C_2
               -A_2 B_1 C_2+A_1 B_2 C_2-A_2 B_2 C_2\bigr)\le 1$}
            & \multicolumn{1}{|c|}{$\mathcal{Q}=\sqrt{2}$} \\
         \hline         
         Number & \multicolumn{1}{|c|}{New Tight Inequalities} & \multicolumn{1}{|c|}{
            Remarks} \\
         \hline
         1 & \multicolumn{1}{|p{9cm}|}{$\mathbb{B}_3^{(++)}\overset{
               A_{1\leftrightarrow 2},C_2\to-C_2}{\longleftrightarrow}
                  \mathbb{B}_3^{(+-)},\ 
               \mathbb{B}_3^{(++)}\overset{A_{1\leftrightarrow 2},
                  A_1\to-A_1,A_2\to-A_2,C_1 \to-C_1}
                     {\longleftrightarrow}\mathbb{B}_3^{(-+)}$,\ 
               $\mathbb{B}_3^{(++)}\overset{A_{1\leftrightarrow 2},
                  B_{1\leftrightarrow 2},A_1\to-A_1,
                     A_2\to-A_2}{\longleftrightarrow}\mathbb{B}_3^{(--)}$} 
            & \multicolumn{1}{|c|}{$\mathcal{Q}=\sqrt{2}$} \\
         \hline                               
         2 & \multicolumn{1}{|p{9cm}|}{$\dfrac{1}{4}\bigl(A_ 2 B_ 1 C_ 1 D_ 1
               -A_ 1 B_ 2 C_ 1 D_ 1-A_ 2 B_ 1 C_ 2 D_ 1+A_ 1 B_ 2 C_ 2 D_ 1
               +A_ 1 B_ 1 C_ 1-A_ 2 B_ 2 C_ 1+A_ 1 B_ 1 C_ 2-A_ 2 B_ 2 C_ 2
               +2 A_ 1 B_ 1 D_ 2+2 A_ 2 B_ 2 D_ 2\bigr)\le 1$} 
            & \multicolumn{1}{|c|}{$\mathcal{Q}=2$} \\
\hline                               
      \end{longtable}
      \begin{longtable}[htbp]{|c|*2{p{2cm}|}}            
         \caption{Two cases of the (4,2,2) tight inequalities generated from 
            $\text{\'Sliwa}_{12}$.} \\
         \hline
         $\text{\'Sliwa}_{12}$ & \multicolumn{1}{|p{9cm}|}{$\dfrac{1}{4}\bigl(2A_1 B_1
               +2A_2 B_2+A_1 C_1+A_2 C_1-B_1 C_1+A_2 B_1 C_1-B_2 C_1-A_1 B_2 C_1
               +A_1 C_2+A_2 C_2-B_1 C_2-A_2 B_1 C_2-B_2 C_2+A_1 B_2 C_2\bigr)\le 1$}
            & \multicolumn{1}{|c|}{$\mathcal{Q}=\sqrt{2}$} \\
         \hline         
         Number & \multicolumn{1}{|c|}{New Tight Inequalities} & \multicolumn{1}{|c|}{
            Remarks} \\
         \hline
         1 & \multicolumn{1}{|c|}{$\mathbb{B}_3^{(++)}\overset{C_1 \to-C_1}
                  {\longleftrightarrow}\mathbb{B}_3^{(+-)},\ 
               \mathbb{B}_3^{(++)}\overset{C_2\to-C_2}{\longleftrightarrow}
                  \mathbb{B}_3^{(-+)}$,\ 
               $\mathbb{B}_3^{(++)}\overset{A\leftrightarrow B}{\longleftrightarrow}
                  \mathbb{B}_3^{(--)}$} 
            & \multicolumn{1}{|c|}{$\mathcal{Q}=\sqrt{2}$} \\
         \hline                               
         2 & \multicolumn{1}{|p{8cm}|}{$\dfrac{1}{4}\bigl(-A_1 B_ 2 C_ 1 D_ 1
               +A_ 2 B_ 1 C_ 1 D_ 2+A_ 1 B_ 2 C_ 2 D_ 1-A_ 2 B_ 1 C_ 2 D_ 2
               +2 A_ 2 B_ 2 D_ 1+2 A_ 1 B_ 1 D_ 2+A_ 2 C_ 1 D_ 1+A_ 1 C_ 1 D_ 2
               +A_ 2 C_ 2 D_ 1+A_ 1 C_ 2 D_ 2-B_ 1 C_ 1-B_ 2 C_ 1-B_ 1 C_ 2-B_ 2 C_ 2
               \bigr)\le 1$} & \multicolumn{1}{|c|}{$\mathcal{Q}=2$} \\
         \hline
         $\mathbb{B}_3^{(+-)}$ & \multicolumn{1}{|p{9cm}|}{$\dfrac{1}{4}\bigl(
               -A_2 B_ 1 C_ 1+A_ 2 B_ 1 C_ 2-A_ 1 B_ 2 C_ 1+A_ 1 B_ 2 C_ 2-2 A_ 1 B_ 1+2 A_ 2 B_ 2-A_ 1 C_ 1+A_ 2 C_ 1-A_ 1 C_ 2+A_ 2 C_ 2-B_ 1 C_ 1
               -B_ 1 C_ 2-B_ 2 C_ 1-B_ 2 C_ 2\bigr)\le 1$} 
            & \multicolumn{1}{|c|}{$\mathbb{B}_3^{(++)}\overset{C_{1\leftrightarrow 2},
               A_1\to-A_1}{\longleftrightarrow}\mathbb{B}_3^{(+-)}$} \\
\hline                               
      \end{longtable}
      \begin{longtable}[htbp]{|c|*2{p{2cm}|}}            
         \caption{Two cases of the (4,2,2) tight inequalities generated from 
            $\text{\'Sliwa}_{13}$.} \\
         \hline
         $\text{\'Sliwa}_{13}$ & \multicolumn{1}{|p{9cm}|}{$\dfrac{1}{4}\bigl(
               2A_1 B_1+2A_2 B_1+A_1 B_1 C_1-A_2 B_1 C_1+A_1 B_2 C_1-A_2 B_2 C_1
               +A_1 B_1 C_2-A_2 B_1 C_2-A_1 B_2 C_2+A_2 B_2 C_2\bigr)\le 1$}
            & \multicolumn{1}{|c|}{$\mathcal{Q}=\sqrt{2}$} \\
         \hline         
         Number & \multicolumn{1}{|c|}{New Tight Inequalities} & \multicolumn{1}{|c|}{
            Remarks} \\
         \hline            
         3 & \multicolumn{1}{|p{8cm}|}{$\dfrac{1}{4}\bigl(A_ 1 B_ 2 C_ 1 D_ 2
               -A_ 2 B_ 2 C_ 1 D_ 2-A_ 1 B_ 2 C_ 2 D_ 2+A_ 2 B_ 2 C_ 2 D_ 2
               +A_ 1 B_ 1 C_ 1-A_ 2 B_ 1 C_ 1+A_ 1 B_ 1 C_ 2-A_ 2 B_ 1 C_ 2
               +A_ 1 B_ 1 D_ 1+A_ 2 B_ 1 D_ 1-A_ 1 B_ 2 D_ 1-A_ 2 B_ 2 D_ 1+A_ 1 B_ 1+A_ 2 B_ 1+A_ 1 B_ 2+A_ 2 B_ 2\bigr)\le 1$} 
            & \multicolumn{1}{|c|}{$\mathcal{Q}=1.58$} \\
         \hline
         $\mathbb{B}_3^{(+-)}$ & \multicolumn{1}{|p{9cm}|}{$\dfrac{1}{4}\bigl(
               A_ 1 B_ 1 C_ 1-A_ 2 B_ 1 C_ 1+A_ 1 B_ 1 C_ 2-A_ 2 B_ 1 C_ 2
               -A_ 1 B_ 2 C_ 1+A_ 2 B_ 2 C_ 1+A_ 1 B_ 2 C_ 2-A_ 2 B_ 2 C_ 2+2 A_ 1 B_ 1
               +2 A_ 2 B_ 1\bigr)\le 1$} 
            & \multicolumn{1}{|c|}{$\mathbb{B}_3^{(++)}\overset{C_{1\leftrightarrow 2}}
               {\longleftrightarrow}\mathbb{B}_3^{(+-)}$} \\
         \hline
         $\mathbb{B}_3^{(-+)}$ & \multicolumn{1}{|p{9cm}|}{$\dfrac{1}{4}\bigl(
               A_ 1 B_ 2 C_ 1-A_ 2 B_ 2 C_ 1-A_ 1 B_ 2 C_ 2+A_ 2 B_ 2 C_ 2
               +A_ 1 B_ 1 C_ 1-A_ 2 B_ 1 C_ 1+A_ 1 B_ 1 C_ 2-A_ 2 B_ 1 C_ 2+2 A_ 1 B_ 2
               +2 A_ 2 B_ 2\bigr)\le 1$} 
            & \multicolumn{1}{|c|}{$\mathbb{B}_3^{(++)}\overset{B_{1\leftrightarrow 2},
               C_2\to-C_2}{\longleftrightarrow}\mathbb{B}_3^{(-+)}$} \\
\hline                                                              
      \end{longtable}
      \begin{longtable}[htbp]{|c|*2{p{2cm}|}}            
         \caption{Two cases of the (4,2,2) tight inequalities generated from 
            $\text{\'Sliwa}_{14}$.} \\
         \hline
         $\text{\'Sliwa}_{14}$ & \multicolumn{1}{|p{9cm}|}{$\dfrac{1}{4}\bigl(
               2A_1 B_1+2A_2 B_1+A_1 C_1-A_2 C_1+A_1 B_2 C_1-A_2 B_2 C_1+A_1 C_2
               -A_2 C_2-A_1 B_2 C_2+A_2 B_2 C_2\bigr)\le 1$} 
            & \multicolumn{1}{|c|}{$\mathcal{Q}=\sqrt{2}$} \\
         \hline         
         Number & \multicolumn{1}{|c|}{New Tight Inequalities} & \multicolumn{1}{|c|}{
            Remarks} \\
         \hline            
         1 & \multicolumn{1}{|p{8cm}|}{$\dfrac{1}{8}\bigl(A_ 1 B_ 1 C_ 1 D_ 1
               -A_ 2 B_ 1 C_ 1 D_ 1+A_ 1 B_ 2 C_ 1 D_ 1-A_ 2 B_ 2 C_ 1 D_ 1
               -A_ 1 B_ 1 C_ 1 D_ 2+A_ 2 B_ 1 C_ 1 D_ 2+A_ 1 B_ 2 C_ 1 D_ 2
               -A_ 2 B_ 2 C_ 1 D_ 2-A_ 1 B_ 1 C_ 2 D_ 1+A_ 2 B_ 1 C_ 2 D_ 1
               -A_ 1 B_ 2 C_ 2 D_ 1+A_ 2 B_ 2 C_ 2 D_ 1+A_ 1 B_ 1 C_ 2 D_ 2
               -A_ 2 B_ 1 C_ 2 D_ 2-A_ 1 B_ 2 C_ 2 D_ 2+A_ 2 B_ 2 C_ 2 D_ 2
               +2 A_ 1 B_ 1 D_ 1+2 A_ 2 B_ 1 D_ 1-2 A_ 1 B_ 2 D_ 1-2 A_ 2 B_ 2 D_ 1
               +2 A_ 1 B_ 1 D_ 2+2 A_ 2 B_ 1 D_ 2+2 A_ 1 B_ 2 D_ 2+2 A_ 2 B_ 2 D_ 2
               +2 A_ 1 C_ 1-2 A_ 2 C_ 1+2 A_ 1 C_ 2-2 A_ 2 C_ 2\bigr)\le 1$} 
            & \multicolumn{1}{|c|}{$\mathcal{Q}=2$} \\
         \hline
         $\mathbb{B}_3^{(+-)}$ & \multicolumn{1}{|p{9cm}|}{$\dfrac{1}{4}\bigl(
               A_ 1 B_ 1 C_ 1-A_ 2 B_ 1 C_ 1-A_ 1 B_ 1 C_ 2+A_ 2 B_ 1 C_ 2-2 A_ 1 B_ 2
               -2 A_ 2 B_ 2+A_ 1 C_ 1-A_ 2 C_ 1+A_ 1 C_ 2-A_ 2 C_ 2\bigr)\le 1$} 
            & \multicolumn{1}{|c|}{$\mathbb{B}_3^{(++)}\overset{B_{1\leftrightarrow 2},
               B_2\to-B_2}{\longleftrightarrow}\mathbb{B}_3^{(+-)}$} \\
\hline                                                              
      \end{longtable} 
      \begin{longtable}[htbp]{|c|*2{p{2cm}|}}            
         \caption{Two cases of the (4,2,2) tight inequalities generated from 
            $\text{\'Sliwa}_{15}$.} \\
         \hline
         $\text{\'Sliwa}_{15}$ & \multicolumn{1}{|p{9cm}|}{$\dfrac{1}{4}\bigl(2A_1 B_1
               +2A_2 B_1+A_1 C_1+A_2 C_1-2B_1 C_1+A_1 B_2 C_1-A_2 B_2 C_1+A_1 C_2
               +A_2 C_2-2B_1 C_2-A_1 B_2 C_2+A_2 B_2 C_2\bigr)\le 1$} 
            & \multicolumn{1}{|c|}{$\mathcal{Q}=1.5$} \\
         \hline         
         Number & \multicolumn{1}{|c|}{New Tight Inequalities} & \multicolumn{1}{|c|}{
            Remarks} \\
         \hline
         1 & \multicolumn{1}{|c|}{$\mathbb{B}_3^{(++)}\overset{A\leftrightarrow C,
                  A_{1\leftrightarrow 2},B_{1\leftrightarrow 2}}{\longleftrightarrow}
                     \mathbb{B}_3^{(+-)},\ 
               \mathbb{B}_3^{(++)}\overset{B_{1\leftrightarrow 2}}{\longleftrightarrow}
                  \mathbb{B}_3^{(-+)}$,\ 
               $\mathbb{B}_3^{(++)}\overset{A\leftrightarrow C,A_{1\leftrightarrow 2}}
                  {\longleftrightarrow}\mathbb{B}_3^{(--)}$} 
            & \multicolumn{1}{|c|}{$\mathcal{Q}=2\sqrt{2}-1$} \\
\hline                                                                         
      \end{longtable}  
      \begin{longtable}[htbp]{|c|*2{p{2cm}|}}            
         \caption{Two cases of the (4,2,2) tight inequalities generated from 
            $\text{\'Sliwa}_{16}$.} \\
         \hline
         $\text{\'Sliwa}_{16}$ & \multicolumn{1}{|p{9cm}|}{$\dfrac{1}{4}\bigl(A_1+A_2
               +A_1 B_1+A_2 B_1+A_1 C_1+A_2 C_1-2A_2 B_1 C_1+A_1 B_2 C_1-A_2 B_2 C_1
               +A_1 B_1 C_2-A_2 B_1 C_2-A_1 B_2 C_2+A_2 B_2 C_2\bigr)\le 1$} 
            & \multicolumn{1}{|c|}{$\mathcal{Q}=1.53$} \\
         \hline         
         Number & \multicolumn{1}{|c|}{New Tight Inequalities} & \multicolumn{1}{|c|}{
            Remarks} \\
         \hline                       
         3 & \multicolumn{1}{|p{9cm}|}{$\dfrac{1}{8}\bigl(-A_1 B_ 1 C_ 1 D_ 1
               -A_ 2 B_ 1 C_ 1 D_ 1+2 A_ 1 B_ 2 C_ 1 D_ 1-2 A_ 2 B_ 2 C_ 1 D_ 1
               -A_ 1 B_ 1 C_ 1 D_ 2-A_ 2 B_ 1 C_ 1 D_ 2-3 A_ 1 B_ 2 C_ 2 D_ 1
               +A_ 2 B_ 2 C_ 2 D_ 1+A_ 1 B_ 2 C_ 2 D_ 2+A_ 2 B_ 2 C_ 2 D_ 2
               +2 A_ 1 B_ 1 C_ 1-2 A_ 2 B_ 1 C_ 1+2 A_ 1 B_ 1 C_ 2-2 A_ 2 B_ 1 C_ 2
               +A_ 1 B_ 1 D_ 1+A_ 2 B_ 1 D_ 1-A_ 1 B_ 2 D_ 1-A_ 2 B_ 2 D_ 1
               +A_ 1 B_ 1 D_ 2+A_ 2 B_ 1 D_ 2+A_ 1 B_ 2 D_ 2+A_ 2 B_ 2 D_ 2
               +A_ 1 C_ 1 D_ 1+A_ 2 C_ 1 D_ 1+A_ 1 C_ 1 D_ 2+A_ 2 C_ 1 D_ 2
               -A_ 1 C_ 2 D_ 1-A_ 2 C_ 2 D_ 1+A_ 1 C_ 2 D_ 2+A_ 2 C_ 2 D_ 2+2 A_ 1 D_ 1
               +2 A_ 2 D_ 1\bigr)\le 1$} 
            & \multicolumn{1}{|c|}{$\mathcal{Q}=\dfrac{5}{3}$} \\
\hline                                                                         
      \end{longtable}
      \begin{longtable}[htbp]{|c|*2{p{2cm}|}}            
         \caption{Two cases of the (4,2,2) tight inequalities generated from 
            $\text{\'Sliwa}_{17}$.} \\
         \hline
         $\text{\'Sliwa}_{17}$ & \multicolumn{1}{|p{9cm}|}{$\dfrac{1}{4}\bigl(A_1+A_2
               +A_1 B_1+A_2 B_1+A_1 C_1+A_2 C_1-A_1 B_1 C_1-A_2 B_1 C_1+2A_1 B_2 C_2
               -2A_2 B_2 C_2\bigr)\le 1$} 
            & \multicolumn{1}{|c|}{$\mathcal{Q}=\sqrt{2}$} \\
         \hline         
         Number & \multicolumn{1}{|c|}{New Tight Inequalities} & \multicolumn{1}{|c|}{
            Remarks} \\
         \hline
         2 & \multicolumn{1}{|c|}{$\mathbb{B}_3^{(++)}\overset{C_{1\leftrightarrow 2}}
                  {\longleftrightarrow}\mathbb{B}_3^{(+-)},\ 
               \mathbb{B}_3^{(++)}\overset{A_{1\leftrightarrow 2},
                  C_{1\leftrightarrow 2},C_2\to-C_2}{\longleftrightarrow}
                     \mathbb{B}_3^{(-+)}$,\ 
               $\mathbb{B}_3^{(++)}\overset{A_{1\leftrightarrow 2},C_1 \to-C_1}
                  {\longleftrightarrow}\mathbb{B}_3^{(--)}$} 
            & \multicolumn{1}{|c|}{$\mathcal{Q}=1.54$} \\
         \hline           
         5 & \multicolumn{1}{|p{8cm}|}{$\dfrac{1}{8}\bigl(-A_1 B_ 1 C_ 1 D_ 1
               -A_ 1 B_ 2 C_ 1 D_ 1-2 A_ 1 B_ 1 C_ 2 D_ 1+2 A_ 1 B_ 2 C_ 2 D_ 1
               -A_ 1 B_ 1 C_ 1 D_ 2+A_ 1 B_ 2 C_ 1 D_ 2+2 A_ 1 B_ 1 C_ 2 D_ 2
               +2 A_ 1 B_ 2 C_ 2 D_ 2-A_ 2 B_ 1 C_ 1 D_ 1-A_ 2 B_ 2 C_ 1 D_ 1
               +2 A_ 2 B_ 1 C_ 2 D_ 1-2 A_ 2 B_ 2 C_ 2 D_ 1-A_ 2 B_ 1 C_ 1 D_ 2
               +A_ 2 B_ 2 C_ 1 D_ 2-2 A_ 2 B_ 1 C_ 2 D_ 2-2 A_ 2 B_ 2 C_ 2 D_ 2
               +A_ 1 B_ 1 D_ 1+A_ 1 B_ 2 D_ 1+A_ 1 B_ 1 D_ 2-A_ 1 B_ 2 D_ 2
               +A_ 2 B_ 1 D_ 1+A_ 2 B_ 2 D_ 1+A_ 2 B_ 1 D_ 2-A_ 2 B_ 2 D_ 2+2 A_ 1 C_ 1
               +2 A_ 2 C_ 1+2 A_ 1+2 A_ 2\bigr)\le 1$} 
            & \multicolumn{1}{|c|}{$\mathcal{Q}=3\sqrt{2}-2$} \\
\hline                                                                         
      \end{longtable} 
      \begin{longtable}[htbp]{|c|*2{p{2cm}|}}            
         \caption{Two cases of the (4,2,2) tight inequalities generated from 
            $\text{\'Sliwa}_{18}$.} \\
         \hline
         $\text{\'Sliwa}_{18}$ & \multicolumn{1}{|p{9cm}|}{$\dfrac{1}{4}\bigl(A_1+A_2
               +A_1 B_1+A_2 B_1+A_1 C_1+A_2 C_1-2B_1 C_1+A_1 B_2 C_1-A_2 B_2 C_1
               +A_1 B_1 C_2-A_2 B_1 C_2+2B_2 C_2-A_1 B_2 C_2-A_2 B_2 C_2\bigr)\le 1$} 
            & \multicolumn{1}{|c|}{$\mathcal{Q}=\dfrac{7-\sqrt{17}}{2}$} \\
         \hline         
         Number & \multicolumn{1}{|c|}{New Tight Inequalities} & \multicolumn{1}{|c|}{
            Remarks} \\
         \hline
         8 & \multicolumn{1}{|p{8cm}|}{$\mathbb{B}_3^{(++)}\overset{
                  A_{1\leftrightarrow 2},B_1\to-B_1,C_1 \to-C_1}
                     {\longleftrightarrow}\mathbb{B}_3^{(+-)}$, \newline
               $\mathbb{B}_3^{(++)}\overset{B_2\to-B_2}
                  {\longleftrightarrow}\mathbb{B}_3^{(-+)}$, \newline
               $\mathbb{B}_3^{(++)}\overset{B_1\to-B_1,C_1 \to-C_1,
                  C_2\to-C_2}{\longleftrightarrow}\mathbb{B}_3^{(--)}$} 
            & \\
         \cline{1-2}           
         11 & \multicolumn{1}{|p{8cm}|}{$\dfrac{1}{4}\bigl(A_ 1 B_ 2 C_ 1 D_ 1
               -A_ 1 B_ 2 C_ 2 D_ 1+A_ 1 B_ 1 C_ 2 D_ 2-A_ 2 B_ 2 C_ 1 D_ 1
               -A_ 2 B_ 2 C_ 2 D_ 1-A_ 2 B_ 1 C_ 2 D_ 2+A_ 1 B_ 1+A_ 2 B_ 1
               +A_ 1 C_ 1 D_ 2+A_ 2 C_ 1 D_ 2+A_ 1+A_ 2+2 B_ 2 C_ 2 D_ 1
               -2 B_ 1 C_ 1 D_ 2\bigr)\le 1$} 
            & \multicolumn{1}{|c|}{$\mathcal{Q}=1.5$} \\
\hline                                                                         
      \end{longtable}
      \begin{longtable}[htbp]{|c|*2{p{2cm}|}}            
         \caption{Two cases of the (4,2,2) tight inequalities generated from 
            $\text{\'Sliwa}_{19}$.} \\
         \hline
         $\text{\'Sliwa}_{19}$ & \multicolumn{1}{|p{9cm}|}{$\dfrac{1}{4}\bigl(A_1+A_2
               +A_1 B_1+A_2 B_1+A_1 C_1+A_2 C_1-2B_1 C_1+2B_2 C_1-A_1 B_2 C_1
               -A_2 B_2 C_1+A_1 B_1 C_2-A_2 B_1 C_2+A_1 B_2 C_2-A_2 B_2 C_2\bigr)
               \le 1$} 
            & \multicolumn{1}{|c|}{$\mathcal{Q}=1.45$} \\
         \hline         
         Number & \multicolumn{1}{|c|}{New Tight Inequalities} & \multicolumn{1}{|c|}{
            Remarks} \\
         \hline                      
         4 & \multicolumn{1}{|p{8cm}|}{$\dfrac{1}{4}\bigl(A_ 1 B_ 2 C_ 1 D_ 1
               -A_ 1 B_ 2 C_ 2 D_ 1+A_ 1 B_ 1 C_ 2 D_ 2-A_ 2 B_ 2 C_ 1 D_ 1
               -A_ 2 B_ 2 C_ 2 D_ 1-A_ 2 B_ 1 C_ 2 D_ 2+A_ 1 B_ 1+A_ 2 B_ 1
               +A_ 1 C_ 1 D_ 2+A_ 2 C_ 1 D_ 2+A_ 1+A_ 2+2 B_ 2 C_ 2 D_ 1
               -2 B_ 1 C_ 1 D_ 2\bigr)\le 1$} & \multicolumn{1}{|c|}{$\mathcal{Q}=1.5$} 
            \\
         \hline
         $\mathbb{B}_3^{(+-)}$ & \multicolumn{1}{|p{9cm}|}{$\dfrac{1}{4}\bigl(
               A_ 1 B_ 2 C_ 1-A_ 1 B_ 1 C_ 2-A_ 1 B_ 2 C_ 2-A_ 2 B_ 2 C_ 1
               +A_ 2 B_ 1 C_ 2-A_ 2 B_ 2 C_ 2+A_ 1 B_ 1+A_ 2 B_ 1-A_ 1 C_ 1-A_ 2 C_ 1
               +A_ 1+A_ 2+2 B_ 1 C_ 1+2 B_ 2 C_ 2\bigr)\le 1$} 
            & \multicolumn{1}{|c|}{$\mathbb{B}_3^{(++)}\overset{B\leftrightarrow C,
               B_{1\leftrightarrow 2},C_{1\leftrightarrow 2}}{\longleftrightarrow}
                  \mathbb{B}_3^{(+-)}$} \\
\hline                                                                         
      \end{longtable}  
      \begin{longtable}[htbp]{|c|*3{p{2cm}|}}            
         \caption{Two cases of the (4,2,2) tight inequalities generated from 
            $\text{\'Sliwa}_{20}$.} \\
         \hline
         $\text{\'Sliwa}_{20}$ & \multicolumn{1}{|p{9cm}|}{$\dfrac{1}{4}\bigl(A_1+A_2
               +A_1 B_1-A_2 B_1+A_1 B_2-A_2 B_2+A_1 C_1-A_2 C_1-B_1 C_1+A_1 B_1 C_1
               +A_2 B_1 C_1-B_2 C_1+A_1 B_2 C_1+A_2 B_2 C_1+B_1 C_2-A_1 B_1 C_2
               -A_2 B_1 C_2-B_2 C_2+A_1 B_2 C_2+A_2 B_2 C_2\bigr)\le 1$} 
            & \multicolumn{1}{|c|}{$\mathcal{Q}=\dfrac{3\sqrt{2}-1}{2}$} \\
         \hline         
         Number & \multicolumn{1}{|c|}{New Tight Inequalities} & \multicolumn{1}{|c|}{
            Remarks} \\
         \hline 
         4 & \multicolumn{1}{|p{8cm}|}{$\mathbb{B}_3^{(++)}\overset{
               A_1\to-A_1}{\longleftrightarrow}\mathbb{B}_3^{(+-)}$,\ 
                  $\mathbb{B}_3^{(++)}\overset{A_2\to-A_2}
                     {\longleftrightarrow}\mathbb{B}_3^{(-+)}$,\ 
            $\mathbb{B}_3^{(++)}\overset{A_1\to-A_1,
               A_2\to-A_2}{\longleftrightarrow}\mathbb{B}_3^{(--)}$} 
            & \multicolumn{1}{|c|}{$\mathcal{Q}=\dfrac{3\sqrt{2}-1}{2}$} \\
         \hline                      
         5 & \multicolumn{1}{|p{8cm}|}{$\dfrac{1}{4}\bigl(A_ 2 B_ 1 C_ 1 D_ 1
               +A_ 2 B_ 2 C_ 1 D_ 1+A_ 1 B_ 1 C_ 1 D_ 2+A_ 1 B_ 2 C_ 1 D_ 2
               -A_ 2 B_ 1 C_ 2 D_ 1+A_ 2 B_ 2 C_ 2 D_ 1-A_ 1 B_ 1 C_ 2 D_ 2
               +A_ 1 B_ 2 C_ 2 D_ 2+A_ 1 B_ 1 D_ 1+A_ 1 B_ 2 D_ 1-A_ 2 B_ 1 D_ 2
               -A_ 2 B_ 2 D_ 2+A_ 1 C_ 1 D_ 1-A_ 2 C_ 1 D_ 2+A_ 2 D_ 1+A_ 1 D_ 2
               -B_ 1 C_ 1-B_ 2 C_ 1+B_ 1 C_ 2-B_ 2 C_ 2\bigr)\le 1$} 
            & \multicolumn{1}{|c|}{$\mathcal{Q}=1.84$} \\
\hline 
      \end{longtable}   
      \begin{longtable}[htbp]{|c|*2{p{2cm}|}}            
         \caption{Two cases of the (4,2,2) tight inequalities generated from 
            $\text{\'Sliwa}_{21}$.} \\
         \hline
         $\text{\'Sliwa}_{21}$ & \multicolumn{1}{|p{9cm}|}{$\dfrac{1}{4}\bigl(A_1+A_2
               +B_1+A_1 B_1+B_2-A_2 B_2+A_1 C_1+A_2 C_1+B_1 C_1-2A_1 B_1 C_1
               -A_2 B_1 C_1+B_2 C_1-A_1 B_2 C_1+A_1 B_1 C_2-A_2 B_1 C_2-A_1 B_2 C_2
               +A_2 B_2 C_2\bigr)\le 1$} 
            & \multicolumn{1}{|c|}{$\mathcal{Q}=1.49$} \\
         \hline         
         Number & \multicolumn{1}{|c|}{New Tight Inequalities} & \multicolumn{1}{|c|}{
            Remarks} \\
         \hline 
         2 & \multicolumn{1}{|p{8cm}|}{$\mathbb{B}_3^{(++)}\overset{
                  C_{1\leftrightarrow 2},C_1 \to-C_1,C_2\to-C_2}
                     {\longleftrightarrow}\mathbb{B}_3^{(+-)}$,\ 
               $\mathbb{B}_3^{(++)}\overset{C_{1\leftrightarrow 2}}{\longleftrightarrow}
                  \mathbb{B}_3^{(-+)}$,\ 
               $\mathbb{B}_3^{(++)}\overset{C_1 \to-C_1,C_2\to-C_2}
                  {\longleftrightarrow}\mathbb{B}_3^{(--)}$} 
            & \multicolumn{1}{|c|}{$\mathcal{Q}=2\sqrt{2}-1$} \\
\hline
      \end{longtable} 
      \begin{longtable}[htbp]{|c|*2{p{2cm}|}}            
         \caption{Two cases of the (4,2,2) tight inequalities generated from 
            $\text{\'Sliwa}_{22}$.} \\
         \hline
         $\text{\'Sliwa}_{22}$ & \multicolumn{1}{|p{9cm}|}{$\dfrac{1}{4}\bigl(A_1+A_2
               +B_1+A_1 B_1+B_2-A_2 B_2+C_1+A_1 C_1+B_1 C_1-2A_1 B_1 C_1-A_2 B_1 C_1
               -A_1 B_2 C_1+A_2 B_2 C_1+C_2-A_2 C_2-A_1 B_1 C_2+A_2 B_1 C_2-B_2 C_2
               +A_1 B_2 C_2\bigr)\le 1$} & \multicolumn{1}{|c|}{$\mathcal{Q}=1.55$} \\
         \hline         
         Number & \multicolumn{1}{|c|}{New Tight Inequalities} & \multicolumn{1}{|c|}{
            Remarks} \\
         \hline 
         2 & \multicolumn{1}{|c|}{$\mathbb{B}_3^{(++)}\overset{C_1 \to-C_1}
                  {\longleftrightarrow}\mathbb{B}_3^{(+-)}$,\ 
               $\mathbb{B}_3^{(++)}\overset{C_2\to-C_2}
                  {\longleftrightarrow}\mathbb{B}_3^{(-+)}$,\ 
               $\mathbb{B}_3^{(++)}\overset{C_1 \to-C_1,C_2\to-C_2}
                  {\longleftrightarrow}\mathbb{B}_3^{(--)}$} 
            & \multicolumn{1}{|c|}{$\mathcal{Q}=1.55$} \\
         \hline                      
         5 & \multicolumn{1}{|p{8cm}|}{$\dfrac{1}{8}\bigl(-A_1 B_ 1 C_ 1 D_ 1
               -2 A_ 1 B_ 2 C_ 1 D_ 1-3 A_ 1 B_ 1 C_ 2 D_ 1-3 A_ 1 B_ 1 C_ 1 D_ 2
               +A_ 1 B_ 1 C_ 2 D_ 2+2 A_ 1 B_ 2 C_ 2 D_ 2-2 A_ 2 B_ 1 C_ 1 D_ 1
               +A_ 2 B_ 2 C_ 1 D_ 1+A_ 2 B_ 2 C_ 2 D_ 1+A_ 2 B_ 2 C_ 1 D_ 2
               +2 A_ 2 B_ 1 C_ 2 D_ 2-A_ 2 B_ 2 C_ 2 D_ 2+2 A_ 1 B_ 1-2 A_ 2 B_ 2
               +A_ 1 C_ 1 D_ 1+A_ 1 C_ 2 D_ 1+A_ 1 C_ 1 D_ 2-A_ 1 C_ 2 D_ 2
               +A_ 2 C_ 1 D_ 1-A_ 2 C_ 2 D_ 1-A_ 2 C_ 1 D_ 2-A_ 2 C_ 2 D_ 2+2 A_ 1
               +2 A_ 2+B_ 1 C_ 1 D_ 1+B_ 2 C_ 1 D_ 1+B_ 1 C_ 2 D_ 1-B_ 2 C_ 2 D_ 1
               +B_ 1 C_ 1 D_ 2-B_ 2 C_ 1 D_ 2-B_ 1 C_ 2 D_ 2-B_ 2 C_ 2 D_ 2+2 B_ 1
               +2 B_ 2+2 C_ 2 D_ 1+2 C_ 1 D_ 2\bigr)\le 1$} 
            & \multicolumn{1}{|c|}{$\mathcal{Q}=2.22$} \\
\hline    
      \end{longtable}   
      \begin{longtable}[htbp]{|c|*2{p{2cm}|}}            
         \caption{Two cases of the (4,2,2) tight inequalities generated from 
            $\text{\'Sliwa}_{23}$.} \\
         \hline
         $\text{\'Sliwa}_{23}$ & \multicolumn{1}{|p{9cm}|}{$\dfrac{1}{4}\bigl(A_1+A_2
               +B_1-A_1 B_1-A_2 B_1+B_2-A_1 B_2-A_2 B_2+A_1 C_1-A_2 C_1-A_1 B_1 C_1
               +A_2 B_1 C_1-A_1 B_2 C_1+A_2 B_2 C_1+B_1 C_2-A_1 B_1 C_2-A_2 B_1 C_2
               -B_2 C_2+A_1 B_2 C_2+A_2 B_2 C_2\bigr)\le 1$} 
            & \multicolumn{1}{|c|}{$\mathcal{Q}=\dfrac{3\sqrt{17}-3}{8}$} \\
         \hline         
         Number & \multicolumn{1}{|c|}{New Tight Inequalities} & \multicolumn{1}{|c|}{
            Remarks} \\
         \hline 
         4 & \multicolumn{1}{|c|}{$\mathbb{B}_3^{(++)}\overset{A_{1\leftrightarrow 2}}
                  {\longleftrightarrow}\mathbb{B}_3^{(+-)}$,\ 
               $\mathbb{B}_3^{(++)}\overset{B_{1\leftrightarrow 2}}{\longleftrightarrow}
                  \mathbb{B}_3^{(-+)}$,\ 
               $\mathbb{B}_3^{(++)}\overset{A_{1\leftrightarrow 2},
                  B_{1\leftrightarrow 2}}{\longleftrightarrow}\mathbb{B}_3^{(--)}$} 
            & \multicolumn{1}{|c|}{$\mathcal{Q}=\dfrac{3\sqrt{17}-3}{8}$} \\
         \hline                      
         5 & \multicolumn{1}{|p{8cm}|}{$\dfrac{1}{4}\bigl(-A_1 B_ 1 C_ 1 D_ 1
               +A_ 1 B_ 2 C_ 2 D_ 1-A_ 1 B_ 2 C_ 1 D_ 2-A_ 1 B_ 1 C_ 2 D_ 2
               +A_ 2 B_ 1 C_ 1 D_ 1+A_ 2 B_ 2 C_ 2 D_ 1+A_ 2 B_ 2 C_ 1 D_ 2
               -A_ 2 B_ 1 C_ 2 D_ 2-A_ 1 B_ 1 D_ 1-A_ 1 B_ 2 D_ 2-A_ 2 B_ 1 D_ 1
               -A_ 2 B_ 2 D_ 2+A_ 1 C_ 1-A_ 2 C_ 1+A_ 1+A_ 2-B_ 2 C_ 2 D_ 1
               +B_ 1 C_ 2 D_ 2+B_ 1 D_ 1+B_ 2 D_ 2\bigr)\le 1$} 
            & \multicolumn{1}{|c|}{$\mathcal{Q}=\dfrac{\sqrt{10}+7}{6}$} \\
\hline 
      \end{longtable}
      \begin{longtable}[htbp]{|c|*2{p{2cm}|}}            
         \caption{Two cases of the (4,2,2) tight inequalities generated from 
            $\text{\'Sliwa}_{24}$.} \\
         \hline
         $\text{\'Sliwa}_{24}$ & \multicolumn{1}{|p{9cm}|}{$\dfrac{1}{5}\bigl(A_1+B_1
               +A_2 B_1+A_1 B_2+A_2 B_2+C_1+A_2 C_1-B_1 C_1+2A_1 B_1 C_1-A_2 B_1 C_1
               -2A_2 B_2 C_1+A_1 C_2+A_2 C_2-2A_2 B_1 C_2-A_1 B_2 C_2+A_2 B_2 C_2
               \bigr)\le 1$} 
            & \multicolumn{1}{|c|}{$\mathcal{Q}=1.588$} \\
         \hline         
         Number & \multicolumn{1}{|c|}{New Tight Inequalities} & \multicolumn{1}{|c|}{
            Remarks} \\
         \hline 
         4 & \multicolumn{1}{|c|}{$\mathbb{B}_3^{(++)}\overset{C_{1\leftrightarrow 2}}
                  {\longleftrightarrow}\mathbb{B}_3^{(+-)}$,\ 
               $\mathbb{B}_3^{(++)}\overset{A\leftrightarrow B}{\longleftrightarrow}
                  \mathbb{B}_3^{(-+)}$,\ 
               $\mathbb{B}_3^{(++)}\overset{A\leftrightarrow B,C_{1\leftrightarrow 2}}
                  {\longleftrightarrow}\mathbb{B}_3^{(--)}$} 
            & \multicolumn{1}{|c|}{$\mathcal{Q}=1.588$} \\
         \hline                      
         11 & \multicolumn{1}{|p{8cm}|}{$\dfrac{1}{5}\bigl(-A_2 B_1 C_1 D_1
               +2 A_1 B_1 C_1 D_2-2 A_2 B_1 C_2 D_2-2 A_2 B_2 C_1 D_1-A_1 B_2 C_2 D_1
               +A_2 B_2 C_2 D_2+A_2 B_1 D_1+A_2 B_2 D_1+A_1 B_2 D_2+A_2 C_1 D_1
               +A_1 C_2 D_1+A_2 C_2 D_2+A_1 D_2-B_1 C_1+B_1+C_1\bigr)\le 1$} 
            & \multicolumn{1}{|c|}{$\mathcal{Q}=\dfrac{9}{5}$} \\
\hline 
      \end{longtable} 
      \begin{longtable}[htbp]{|c|*2{p{2cm}|}}            
         \caption{Two cases of the (4,2,2) tight inequalities generated from 
            $\text{\'Sliwa}_{25}$.} \\
         \hline
         $\text{\'Sliwa}_{25}$ & \multicolumn{1}{|p{9cm}|}{$\dfrac{1}{5}\bigl(A_1+B_1
               +A_2 B_1+A_1 B_2+A_2 B_2+C_1+A_2 C_1-B_1 C_1+2A_1 B_1 C_1-A_2 B_1 C_1
               -2A_2 B_2 C_1+A_1 C_2+A_2 C_2-2A_1 B_1 C_2+A_1 B_2 C_2-A_2 B_2 C_2
               \bigr)\le 1$} 
            & \multicolumn{1}{|c|}{$\mathcal{Q}=1.36$} \\
         \hline         
         Number & \multicolumn{1}{|c|}{New Tight Inequalities} & \multicolumn{1}{|c|}{
            Remarks} \\
         \hline                                  
         2 & \multicolumn{1}{|p{8cm}|}{$\dfrac{1}{10}\bigl(2 A_ 1 B_ 1 C_ 1 D_ 1
               +2 A_ 1 B_ 2 C_ 1 D_ 1-A_ 1 B_ 1 C_ 2 D_ 1-A_ 1 B_ 2 C_ 2 D_ 1
               +2 A_ 1 B_ 1 C_ 1 D_ 2-2 A_ 1 B_ 2 C_ 1 D_ 2-3 A_ 1 B_ 1 C_ 2 D_ 2
               +3 A_ 1 B_ 2 C_ 2 D_ 2-3 A_ 2 B_ 1 C_ 1 D_ 1-3 A_ 2 B_ 2 C_ 1 D_ 1
               -A_ 2 B_ 1 C_ 2 D_ 1-A_ 2 B_ 2 C_ 2 D_ 1+A_ 2 B_ 1 C_ 1 D_ 2
               -A_ 2 B_ 2 C_ 1 D_ 2+A_ 2 B_ 1 C_ 2 D_ 2-A_ 2 B_ 2 C_ 2 D_ 2
               +A_ 1 B_ 1 D_ 1+A_ 1 B_ 2 D_ 1-A_ 1 B_ 1 D_ 2+A_ 1 B_ 2 D_ 2
               +2 A_ 2 B_ 1 D_ 1+2 A_ 2 B_ 2 D_ 1+2 A_ 1 C_ 2+2 A_ 2 C_ 1+2 A_ 2 C_ 2
               +2 A_ 1-B_ 1 C_ 1 D_ 1-B_ 2 C_ 1 D_ 1-B_ 1 C_ 1 D_ 2+B_ 2 C_ 1 D_ 2
               +B_ 1 D_ 1+B_ 2 D_ 1+B_ 1 D_ 2-B_ 2 D_ 2+2 C_ 1\bigr)\le 1$} 
            & \multicolumn{1}{|c|}{$\mathcal{Q}=\dfrac{8\sqrt{2}-3}{5}$} \\
\hline  
      \end{longtable}
      \begin{longtable}[htbp]{|c|*2{p{2cm}|}}            
         \caption{Two cases of the (4,2,2) tight inequalities generated from 
            $\text{\'Sliwa}_{26}$.}\label{tab:Tigh4PSliwa26} \\
         \hline
         $\text{\'Sliwa}_{26}$ & \multicolumn{1}{|p{9cm}|}{$\dfrac{1}{5}\bigl(A_1+B_1
               +A_1 B_1+2A_2 B_2+C_1+A_1 C_1+B_1 C_1-A_1 B_1 C_1-2A_2 B_2 C_1
               +2A_2 C_2-2A_2 B_1 C_2-2B_2 C_2+2A_1 B_2 C_2\bigr)\le 1$} 
            & \multicolumn{1}{|c|}{$\mathcal{Q}=\dfrac{4\sqrt{3}+1}{5}$} \\
         \hline         
         Number & \multicolumn{1}{|c|}{New Tight Inequalities} & \multicolumn{1}{|c|}{
            Remarks} \\
         \hline
         1 & \multicolumn{1}{|c|}{$\mathbb{B}_3^{(++)}\overset{B_1\to-B_1}
               {\longleftrightarrow}\mathbb{B}_3^{(+-)}$,\ 
            $\mathbb{B}_3^{(++)}\overset{A\leftrightarrow C}{\longleftrightarrow}
               \mathbb{B}_3^{(-+)}$,\ 
            $\mathbb{B}_3^{(++)}\overset{A\leftrightarrow C,B_1\to-B_1}
               {\longleftrightarrow}\mathbb{B}_3^{(--)}$} 
         & \multicolumn{1}{|c|}{$\mathcal{Q}=\dfrac{4\sqrt{3}+1}{5}$} \\
\hline
      \end{longtable}
      \begin{longtable}[htbp]{|c|*2{p{2cm}|}}            
         \caption{Two cases of the (4,2,2) tight inequalities generated from 
            $\text{\'Sliwa}_{27}$.} \\
         \hline
         $\text{\'Sliwa}_{27}$ & \multicolumn{1}{|p{11cm}|}{$\dfrac{1}{5}\bigl(2A_1
               +A_2+B_1-A_1 B_1+A_1 B_2+A_2 B_2+C_1-A_1 C_1+2A_1 B_1 C_1-2A_2 B_1 C_1
               +B_2 C_1-A_1 B_2 C_1+A_1 C_2+A_2 C_2+B_1 C_2-A_1 B_1 C_2+B_2 C_2
               -2A_1 B_2 C_2-A_2 B_2 C_2\bigr)\le 1$} 
            & \multicolumn{1}{|c|}{$\mathcal{Q}=1.39$} \\
         \hline         
         Number & \multicolumn{1}{|c|}{New Tight Inequalities} & \multicolumn{1}{|c|}{
            Remarks} \\
         \hline
         5 & \multicolumn{1}{|c|}{$\mathbb{B}_3^{(++)}\overset{C_{1\leftrightarrow 2}}
               {\longleftrightarrow}\mathbb{B}_3^{(+-)}$,\ 
            $\mathbb{B}_3^{(++)}\overset{C_{1\leftrightarrow 2},C_1 \to-C_1,
               C_2\to-C_2}{\longleftrightarrow}\mathbb{B}_3^{(-+)}$,\ 
            $\mathbb{B}_3^{(++)}\overset{C_1 \to-C_1,C_2\to-C_2}
               {\longleftrightarrow}\mathbb{B}_3^{(--)}$} & \\
\hline
      \end{longtable}%
   \subsection{Some $(5,2,2)$ Inequalities}
      In this subsection we will start from the third $(4,2,2)$ inequality shown in Table~\ref{tab:Tigh4PSliwa3}, with the help of equivalent transformations and iterative formula 
      \begin{align}
         \mathbb{B}_5 =\dfrac{1}{2} \Bigl[\mathbb{B}_4^{(++)} (E_1 +E_2) 
            +\mathbb{B}_4^{(+-)} (1-E_2) +\mathbb{B}_4^{(-+)} (1-E_1)\Bigl]\le 1,
      \end{align}
      under the condition
      \begin{equation}
         \mathbb{B}_4^{(--)} =\mathbb{B}_4^{(+-)} +\mathbb{B}_4^{(-+)} 
            -\mathbb{B}_4^{(++)},
      \end{equation}
      to build a family of tight $(5,2,2)$ Bell inequalities. Akin to the circumstance of last subsection, the $\mathbb{B}_4^{(++)}$ is set to the third $(4,2,2)$ inequality in Table~\ref{tab:Tigh4PSliwa3}, which is listed in the first row of the following table. Note that all newly generated inequalities are normalized and $Q$ refers to the numerical quantum upper bound of corresponding inequality. Likewise, the newly established $(5,2,2)$ inequalities are too cumbersome to be presented completely, so we print two of them merely and list them entirely in \cite{NewTightBI}.
      \begin{longtable}[htbp]{|c|*2{p{2cm}|}}            
         \caption{Two cases of the (5,2,2) tight inequalities generated from the third 
            $(4,2,2)$ inequality in Table~\ref{tab:Tigh4PSliwa3}.} \\
         \hline
         \multicolumn{1}{|p{2cm}|}{The $(4,2,2)$ Inequality} 
            & \multicolumn{1}{|p{11cm}|}{$\dfrac{1}{4}\bigl(A_ 1 B_ 1 C_ 1 D_ 1
                  +A_ 2 B_ 1 C_ 1 D_ 1+A_ 1 B_ 2 C_ 1 D_ 1+A_ 2 B_ 2 C_ 1 D_ 1
                  +A_ 2 B_ 1 C_ 1 D_ 2-A_ 1 B_ 2 C_ 1 D_ 2-A_ 1 B_ 1 C_ 2 D_1
                  +A_ 2 B_ 1 C_ 2 D_ 1+A_ 1 B_ 2 C_ 2 D_ 1-A_ 2 B_ 2 C_ 2 D_ 1
                  -A_ 2 B_1 C_ 2 D_ 2+A_ 1 B_ 2 C_ 2 D_ 2+A_ 1 B_ 1 C_ 1
                  -A_ 2 B_ 2 C_ 1+A_ 1 B_1 C_ 2-A_ 2 B_ 2 C_ 2\bigr)\le 1$} 
               & \multicolumn{1}{|c|}{$\mathcal{Q}=2$} \\
         \hline         
         Number & \multicolumn{1}{|c|}{New Tight Inequalities} 
            & \multicolumn{1}{|c|}{Remarks} \\
         \hline           
         1 & \multicolumn{1}{|p{11cm}|}{$\dfrac{1}{8}\bigl(2 A_ 1 B_ 1 C_ 1 D_ 1 E_ 1
               +2 A_ 1 B_ 2 C_ 1 D_ 1 E_ 1-2 A_ 1 B_ 2 C_ 1 D_ 2 E_ 1
               +2 A_ 2 B_ 2 C_ 1 D_ 1 E_ 2+A_ 2 B_ 1 C_ 1 D_ 2 E_ 2
               +A_ 2 B_ 2 C_ 1 D_ 2 E_ 2+2 A_ 2 B_ 1 C_ 2 D_ 1 E_ 1
               -2 A_ 2 B_ 2 C_ 2 D_ 1 E_ 1-2 A_ 2 B_ 1 C_ 2 D_ 2 E_ 1
               +2 A_ 1 B_ 2 C_ 2 D_ 1 E_ 2+A_ 1 B_ 1 C_ 2 D_ 2 E_ 2
               +A_ 1 B_ 2 C_ 2 D_ 2 E_ 2+2 A_ 2 B_ 1 C_ 1 D_ 1+A_ 2 B_ 1 C_ 1 D_ 2
               -A_ 2 B_ 2 C_ 1 D_ 2-2 A_ 1 B_ 1 C_ 2 D_ 1-A_ 1 B_ 1 C_ 2 D_ 2
               +A_ 1 B_ 2 C_ 2 D_ 2+2 A_ 1 B_ 1 C_ 1 E_ 1+A_ 2 B_ 1 C_ 1 E_ 2
               -A_ 2 B_ 2 C_ 1 E_ 2-2 A_ 2 B_ 2 C_ 2 E_ 1+A_ 1 B_ 1 C_ 2 E_ 2
               -A_ 1 B_ 2 C_ 2 E_ 2-A_ 2 B_ 1 C_ 1-A_ 2 B_ 2 C_ 1+A_ 1 B_ 1 C_ 2
               +A_ 1 B_ 2 C_ 2\bigr)\le 1$} 
            & \multicolumn{1}{|c|}{$\mathcal{Q}=2.01$} \\
         \hline
         $\mathbb{B}_4^{(+-)}$ & \multicolumn{1}{|p{11cm}|}{$\dfrac{1}{4}\bigl(
               A_ 1 B_ 1 C_ 1 D_ 1+A_ 2 B_ 1 C_ 1 D_ 1+A_ 1 B_ 2 C_ 1 D_ 1
               -A_ 2 B_ 2 C_ 1 D_ 1-A_ 1 B_ 2 C_ 1 D_ 2-A_ 2 B_ 2 C_ 1 D_ 2
               -A_ 1 B_ 1 C_ 2 D_ 1+A_ 2 B_ 1 C_ 2 D_ 1-A_ 1 B_ 2 C_ 2 D_ 1
               -A_ 2 B_ 2 C_ 2 D_ 1-A_ 1 B_ 1 C_ 2 D_ 2-A_ 2 B_ 1 C_ 2 D_ 2
               +A_ 1 B_ 1 C_ 1-A_ 2 B_ 1 C_ 1+A_ 1 B_ 2 C_ 2-A_ 2 B_ 2 C_ 2
               \bigr)\le 1$} 
            & \multicolumn{1}{|c|}{$\mathbb{B}_4^{(++)}\overset{A\leftrightarrow C,A_1\to-A_1,
               C_1 \to-C_1}{\longleftrightarrow}\mathbb{B}_4^{(+-)}$} \\
         \hline
         $\mathbb{B}_4^{(-+)}$ & \multicolumn{1}{|p{11cm}|}{$\dfrac{1}{4}\bigl(
               -A_1 B_ 1 C_ 1 D_ 1+A_ 2 B_ 1 C_ 1 D_ 1-A_ 1 B_ 2 C_ 1 D_ 1
               +A_ 2 B_ 2 C_ 1 D_ 1+A_ 2 B_ 1 C_ 1 D_ 2+A_ 1 B_ 2 C_ 1 D_ 2
               -A_ 1 B_ 1 C_ 2 D_ 1-A_ 2 B_ 1 C_ 2 D_ 1+A_ 1 B_ 2 C_ 2 D_ 1
               +A_ 2 B_ 2 C_ 2 D_ 1+A_ 2 B_ 1 C_ 2 D_ 2+A_ 1 B_ 2 C_ 2 D_ 2
               -A_ 1 B_ 1 C_ 1-A_ 2 B_ 2 C_ 1+A_ 1 B_ 1 C_ 2+A_ 2 B_ 2 C_ 2
               \bigr)\le 1$} 
            & \multicolumn{1}{|c|}{$\mathbb{B}_4^{(++)}\overset{A_1\to-A_1,C_2\to-C_2}
               {\longleftrightarrow}\mathbb{B}_4^{(-+)}$} \\
         \hline
         $\mathbb{B}_4^{(--)}$ & \multicolumn{1}{|p{11cm}|}{$\dfrac{1}{4}\bigl(
               -A_1 B_ 1 C_ 1 D_ 1+A_ 2 B_ 1 C_ 1 D_ 1-A_ 1 B_ 2 C_ 1 D_ 1
               -A_ 2 B_ 2 C_ 1 D_ 1+A_ 1 B_ 2 C_ 1 D_ 2-A_ 2 B_ 2 C_ 1 D_ 2
               -A_ 1 B_ 1 C_ 2 D_ 1-A_ 2 B_ 1 C_ 2 D_ 1-A_ 1 B_ 2 C_ 2 D_ 1
               +A_ 2 B_ 2 C_ 2 D_ 1-A_ 1 B_ 1 C_ 2 D_ 2+A_ 2 B_ 1 C_ 2 D_ 2
               -A_ 1 B_ 1 C_ 1-A_ 2 B_ 1 C_ 1+A_ 1 B_ 2 C_ 2+A_ 2 B_ 2 C_ 2
               \bigr)\le 1$} 
            & \multicolumn{1}{|c|}{$\mathbb{B}_4^{(++)}\overset{A\leftrightarrow C,
               A_1\to-A_1,A_2\to-A_2}{\longleftrightarrow}\mathbb{B}_4^{(--)}$} \\                   
         \hline
         117 & \multicolumn{1}{|c|}{$\mathbb{B}_4^{(++)}\overset{C_{1\leftrightarrow 2},
                  D_1\to-D_1}{\longleftrightarrow}\mathbb{B}_4^{(+-)}$,\ 
               $\mathbb{B}_4^{(++)}\overset{A\leftrightarrow B,C_{1\leftrightarrow 2}}
                  {\longleftrightarrow}\mathbb{B}_4^{(-+)}$,\ 
               $\mathbb{B}_4^{(++)}\overset{A\leftrightarrow B,D_1\to-D_1}{\longleftrightarrow}
                  \mathbb{B}_4^{(--)}$} 
            & \multicolumn{1}{|c|}{$\mathcal{Q}=2$} \\
         \hline                                                             
      \end{longtable}
   \bibliography{sup}